# Quantifying the Distribution of Biexciton Emission Efficiencies in Colloidal Quantum Shells

Tjom Arens[1], Dulanjan Harankahage[2,3], Divesh Nazar[2,3], Mikhail Zamkov[2,3], and Freddy T. Rabouw[1,*]

[1] Debye Institute for Nanomaterials Science, Utrecht University, Princetonplein 1, 3584CC Utrecht, The Netherlands
[2] The Center for Photochemical Sciences, Bowling Green State University, Bowling Green, Ohio 43403, USA
[3] Department of Physics, Bowling Green State University, Bowling Green, Ohio 43403, USA
*Corresponding author: f.t.rabouw@uu.nl

**Abstract**

The efficiency of multi-photon emission is an important characteristic of quantum light sources. Bright multi-photon emission is desirable for high-power lighting and lasers, while its complete suppression is required for high-purity single-photon generation. In colloidal quantum emitters, multi-photon emission can vary significantly between individual particles. Resolving this heterogeneity remains challenging with conventional particle-by-particle approaches.

Here, we introduce a crosstalk-suppressed SPAD-array photon-correlation approach for high-throughput quantification of multi-photon emission from more than 1000 colloidal quantum shells. By projecting two images of the same sample onto distant regions of the detector array, we avoid short-range crosstalk between detector pixels. Time gating suppresses dark-count coincidences and distinguishes individual emitters from clusters. Applying this method to quantum shells reveals a near-Gaussian distribution of biexciton emission efficiencies, with a mean of 0.55 and an estimated intrinsic standard deviation of 0.12. Intra-batch correlations between the biexciton efficiency and the particle brightness are consistent with the volume scaling of Auger quenching. These results establish SPAD-array photon correlation as a scalable route to resolve multi-photon heterogeneities in nanoparticle ensembles.

# Introduction

Colloidal semiconductor nanocrystals are a versatile platform for tailoring photon statistics, ranging from anti-bunched single-photon emission to efficient multiexciton emission. Strongly confined colloidal quantum dots offer high single-photon purity, as fast Auger recombination efficiently quenches multiexciton states.[1–3] In contrast, heterostructures such as thick-shell quantum dots,[4,5] dot-in-rods,[6] and quantum shells[7,8] promote multiexciton emission[9] by suppressing carrier–carrier interactions and thereby reducing Auger recombination.

Chemically synthesized nanocrystal batches inevitably exhibit particle-to-particle variations in size, shape, shell thickness, core–shell interface, and composition. These structural differences translate into a distribution of biexciton efficiencies across the ensemble,[1,10–13] as nonradiative Auger recombination is highly sensitive to nanocrystal structure.[14,15] Ensemble-averaged photon statistics reveal only average biexciton efficiency,[16–18] while assessing the distribution requires single-particle photon-correlation measurements. However, single-particle measurements are time consuming, limiting sample size and statistical robustness of interpretations. Quantifying batch-level biexciton-efficiency heterogeneity would require methods that retain single-particle resolution while scaling to statistically meaningful numbers of emitters.

Single-photon avalanche diode (SPAD) arrays provide a route toward such high-throughput photon-correlation measurements.[19,20] These detectors combine spatially resolved imaging with single-photon sensitivity and are available in formats ranging from smaller arrays to devices with hundreds of thousands of pixels.[21–24] Photon-correlation measurements on many emitters in parallel would reduce acquisition times and limit systematic bias because all emitters are probed under identical experimental conditions.

In practice, SPAD-array correlation measurements face technical challenges. A key problem is optical and electrical crosstalk: a detection event in one pixel can induce a spurious detection event in a neighboring pixel, producing artificial photon coincidences.[25] Such coincidences are particularly problematic for measurements of multi-exciton emission, which rely on the zero-delay second-order correlation, $g^{(2)}(0)$, of the emission of an individual particle. Artificial coincidences can be misinterpreted as genuine photon pairs and lead to an overestimation of the biexciton efficiency.[26] The first demonstration of SPAD-array-based high-throughput photon-correlation[19] addressed crosstalk by calibrating the expected magnitude and subtracting it from the measured correlation signal. While this correction reduces systematic offsets in $g^{(2)}(0)$, it remains detector- and condition-dependent, requiring recalibration under varying operating conditions such as temperature, humidity and bias voltage. Moreover, subtraction removes only the mean crosstalk contribution, while the associated shot noise remains. Dark counts present a second challenge.[20,26] Background subtraction of dark-count coincidences is possible, but again does not remove shot noise. Experiments at higher excitation power can improve the signal-to-background ratio, but also perturb the multiexciton population.[1] The finite spatial resolution of optical microscopy presents a third challenge. A fluorescence spot of quantum emitters on a substrate may contain a small cluster of emitters, unresolved by the diffraction-limited spatial resolution of even the best microscope. The $g^{(2)}(0)$ measured on a single fluorescence spot may thus include significant cluster-induced coincidences,[27] distorting the characterization of intrinsic single-particle properties.

Here, we address all three challenges in a single workflow and present high-throughput characterization of multi-photon emission by quantum emitters. The SPAD array is operated in a frame-based mode, with one detector frame synchronized to each excitation pulse. Two photon detections from the same emitter within one frame therefore signal a coincidence event. We avoid cross-talk entirely by projecting two identical images of the same sample onto distinct regions of the SPAD array and cross-correlating the two images.[28] Time gating reduces dark counts by a factor 50, so that they become negligible compared to fluorescence counts. Time gating also allows for reference measurements that distinguish individual emitters from clusters.[27,29]

We apply this combined approach to quantum shells, spherical quantum wells in which the emissive layer is sandwiched between wider-gap inner and outer shells.[7] This architecture suppresses Auger recombination and is therefore a strong candidate for biexciton-based applications. We analyze measurements on more than 1000 quantum shells, extracting that the single-particle distribution of biexciton emission efficiencies is approximately Gaussian, with a mean of 0.55 and an estimated intrinsic standard deviation of 0.12. Our results demonstrate both the statistical power of the SPAD-array multi-photon imaging and the potential of quantum shells for optical-gain applications.

# Results & Discussion

## Distribution of biexciton efficiencies determines ensemble gain

The multi-photon characteristics of an ensemble of quantum emitters is often discussed in terms of an average biexciton emission efficiency $\eta_{\mathrm{BX}}$ (Figure 1a),[7,16,17] i.e. the probability that a biexciton decays radiatively (rate constant $k_{\mathrm{rad,BX}}$) rather than through nonradiative Auger recombination (rate constant $k_{\mathrm{Auger,BX}}$):

$$\eta_{\mathrm{BX}} = \frac{k_{\mathrm{rad,BX}}}{k_{\mathrm{rad,BX}} + k_{\mathrm{Auger,BX}}}. \tag{1}$$

$\eta_{\mathrm{BX}}$ is particularly important under high-excitation conditions, where it determines how efficiently emitters contribute to optical gain and lasing. However, the *average* biexciton efficiency $\langle \eta_{\mathrm{BX}} \rangle$ does not uniquely determine the gain behavior of the ensemble, because particle-to-particle variations can have a significant influence.

Figure 1 illustrates why the full distribution of $\eta_{\mathrm{BX}}$ matters. We compare two limiting scenarios with identical average biexciton efficiency, $\langle \eta_{\mathrm{BX}} \rangle$, but different single-particle distributions (Figure 1b). In scenario A, all emitters have the same $\eta_{\mathrm{BX}}$. In scenario B, the ensemble consists of two subpopulations: one with $\eta_{\mathrm{BX}} = 1$ and one with $\eta_{\mathrm{BX}} = 0$. Although the two ensembles have the same $\langle \eta_{\mathrm{BX}} \rangle$, their gain behavior is fundamentally different. Under continuous-wave excitation, the calculated steady-state absorption evolves differently with excitation rate for the two scenarios (Figure 1c). For scenario B the threshold excitation rate to reach gain is lower, but the gain saturates at incomplete inversion; scenario A has a higher threshold but stronger gain at high excitation rates. The difference between the scenarios is even more pronounced for pulsed excitation, where the calculated gain lifetime for an ensemble with moderate $\langle \eta_{\mathrm{BX}} \rangle$ is more than an order of magnitude longer for scenario A than for scenario B (Figure 1d). These calculations motivate statistically robust single-particle measurements that can quantify the full distribution of multi-photon emission efficiencies.

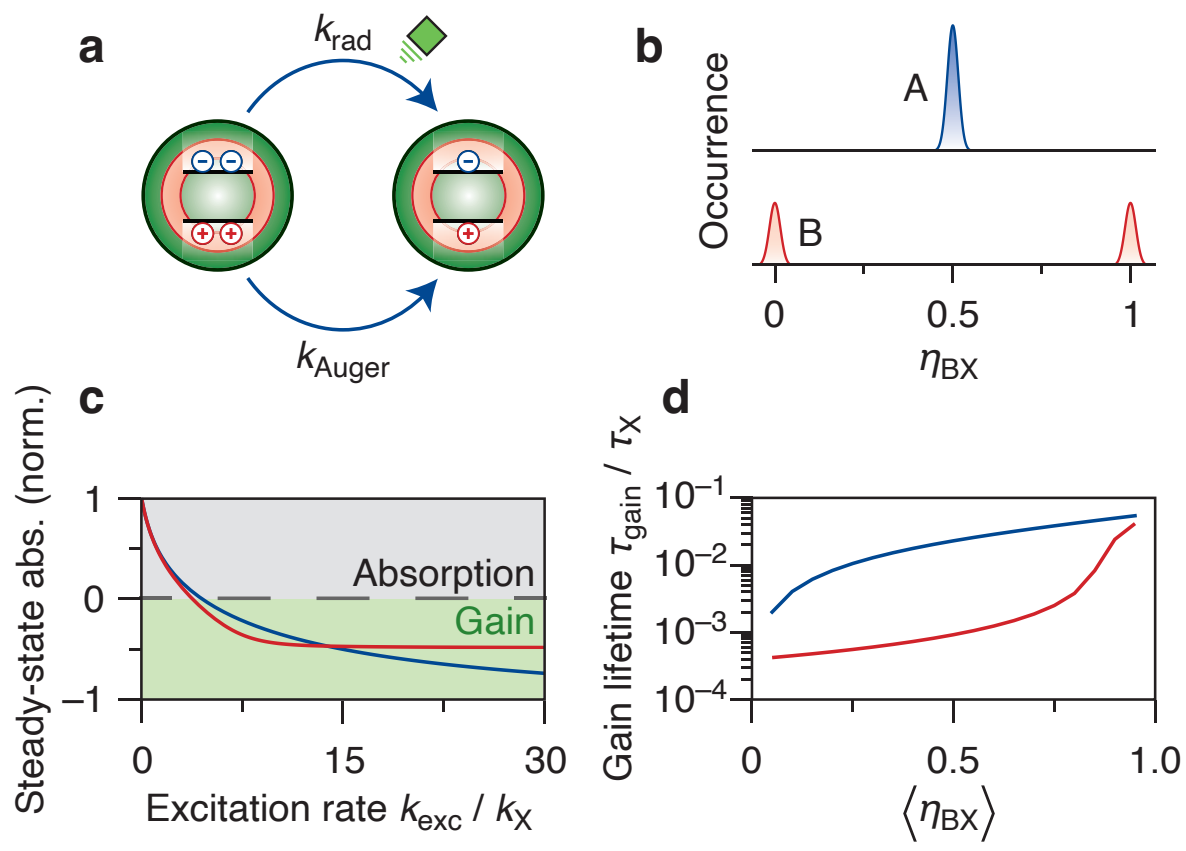

**Figure 1. Effect of the biexciton-efficiency distribution on ensemble gain.** (a) The biexciton emission efficiency, $\eta_{BX}$, describes the probability that a biexciton decays radiatively rather than through nonradiative Auger recombination: $\eta_{BX} = k_{rad,BX}/(k_{rad,BX} + k_{Auger,BX})$. (b) Two ensembles with identical $\langle\eta_{BX}\rangle = 0.5$: a homogeneous ensemble in which all emitters have $\eta_{BX} = 0.5$, and a heterogeneous ensemble consisting of equal fractions of emitters with $\eta_{BX} = 1$ and $\eta_{BX} = 0$. (c) Calculated steady-state absorption under continuous-wave excitation as a function of excitation rate $k_{exc}$ (normalized to the exciton decay rate $k_X$) for scenario A (blue) and B (red). (d) Calculated gain lifetime $\tau_{gain}$ (normalized to the exciton lifetime $\tau_X$) under pulsed excitation, with average number of excitations per particle of $\mu = 1.7$, as a function of $\langle\eta_{BX}\rangle$. Although the two ensembles have the same average biexciton efficiency, their gain thresholds, saturation behavior, and gain lifetimes differ substantially. The calculations consider statistical scaling of rate constants (Supporting Information Section S2).[30]

## Crosstalk-free SPAD-array measurements

SPAD arrays are a promising technology for high-throughput single-photon experiments, but they face technical challenges. The first challenge is crosstalk between pixels (Figure 2a). A detection event in one pixel can trigger a spurious detection event in a neighboring pixel, producing artificial coincidences that are indistinguishable from true photon pairs if both detections are assigned to the same emitter. These false coincidences contribute to $g^{(2)}(0)$ and thereby directly lead to an overestimation of the biexciton emission efficiency.[26]

We quantify the spatial dependence of crosstalk on our SPAD array by illuminating all pixels uniformly with an uncorrelated light source. The measured crosstalk probability is highest for neighboring pixels and decreases exponentially with distance (Figure 2b; Supporting Information Section S3).

Based on this strong distance dependence, we implement the multi-emitter photon-correlation scheme shown in Figure 2c. A sample of quantum shells on a glass slide is widefield illuminated, exciting tens to hundreds of quantum shells simultaneously. The emitted light is split by a 50:50 beamsplitter and projected onto two spatially separated regions of the SPAD array. Each emitter therefore appears twice on the detector, once in each region. The corresponding images are separated by more than 200 pixels, reducing the crosstalk probability to effectively zero. We can thus calculate photon correlations between the two corresponding images of the same emitter without crosstalk artifacts.

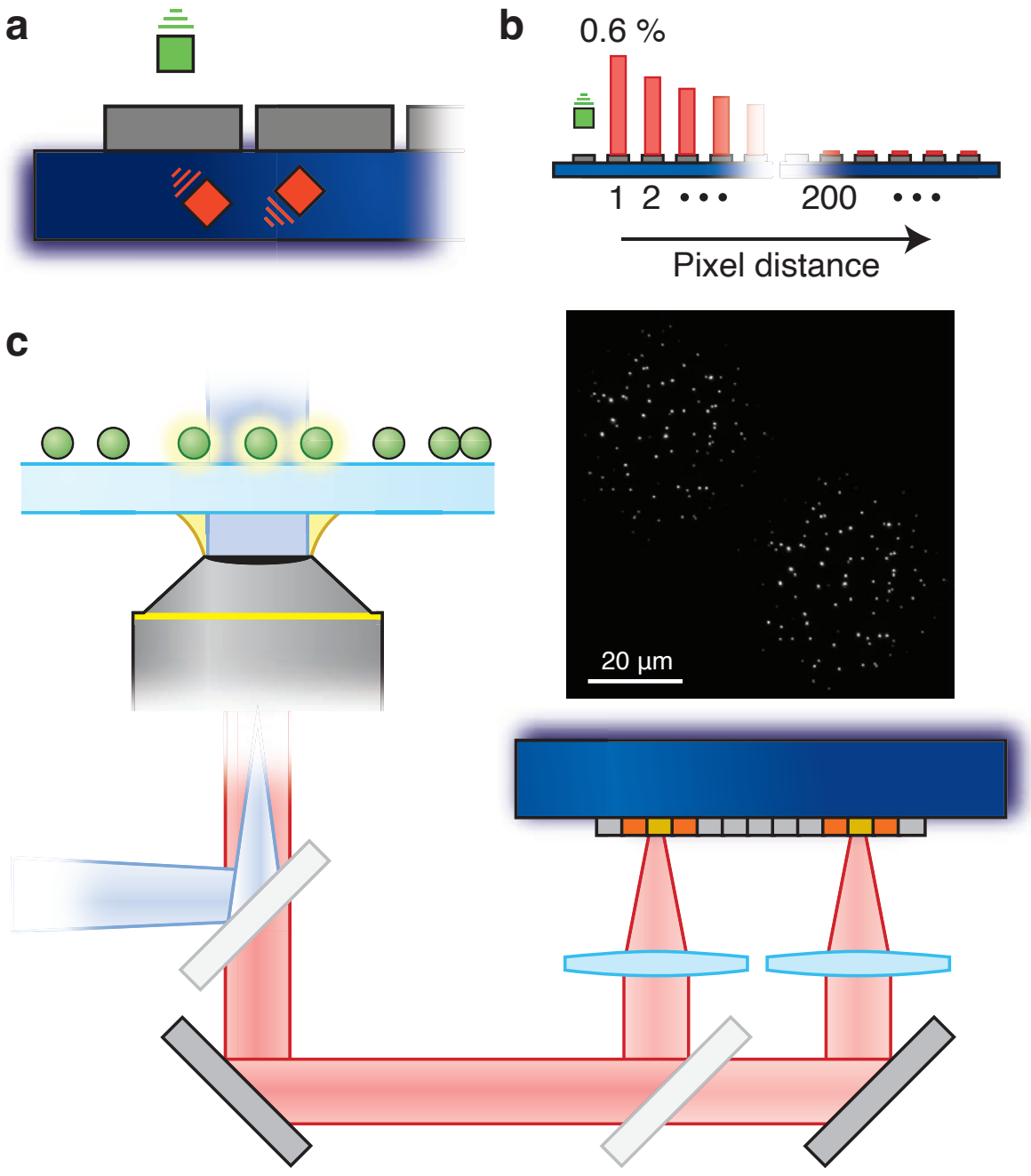

**Figure 2. Spatial separation suppresses detector crosstalk.** (a) Schematic of detector crosstalk, where the detection avalanche in one pixel generates a photon that is collected by a neighboring pixel and induces a spurious detection event there. (b) Measured crosstalk probability as a function of pixel separation, showing a rapid decrease with distance. (c) Experimental detection scheme. Emission from the same sample is split onto two spatially separated regions of the SPAD array, enabling cross-correlations between corresponding emitter images at separations where short-range crosstalk is negligible.

## Frame-based $g^{(2)}(0)$ measurements

To determine $g^{(2)}(0)$ for each emission spot, we adapt the principle of a pulsed Hanbury-Brown–Twiss measurement to a frame-based SPAD-array detector. In a conventional two-detector experiment, photons emitted are split onto two SPAD detectors. Upon pulsed excitation, detection of one photon on each detector within the same excitation cycle contributes to the zero-delay coincidence peak. Coincidences between photons detected in different excitation cycles provide the reference side peaks. The ratio between the zero-delay peak and the side peaks yields the normalized $g^{(2)}(0)$, which is equal to $\eta_{\mathrm{BX}}$ under low-excitation conditions and considering unity exciton efficiency.[1,31]

In our experiment, the emitters are quantum shells: spherical quantum-well nanocrystals designed to suppress Auger recombination and thereby support efficient biexciton emission.[7,8,16] For each diffraction-limited quantum-shell spot, the goal is to determine how often a single excitation pulse produces one detected photon and how often it produces a detected photon pair. Whereas conventional single-pixel SPADs provide a precise time stamp (sub-nanosecond resolution) on each individual photon, the SPAD512 detector does not. Each of the $512 \times 512$ pixels offers single-photon sensitivity, but no time stamps. Instead, a camera frame records the number of photons detected on each pixel during a defined integration window.

We synchronize the experiment such that each camera frame measures the emission following one laser pulse (Figure 3a). A pulse generator starts the camera integration and also triggers the laser. Each camera frame therefore contains the photon detections associated with a single excitation cycle. In practice, the maximum frame rate of the camera is 80 kHz, so we choose this as our repetition period. The camera frames have a 1-bit depth at these highest frame rates, i.e. each pixel records zero or one photon. This is suitable for our purpose of photon and photon-pair counting.

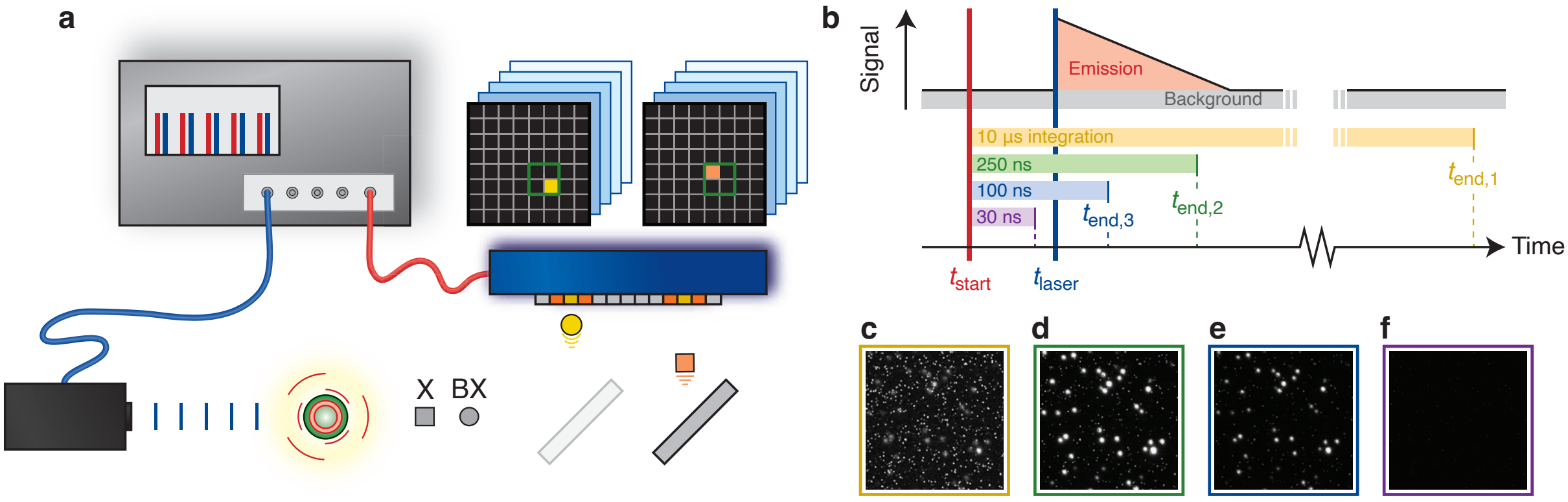

**Figure 3. Frame-based $g^{(2)}(0)$ measurements.** (a) Synchronization scheme for frame-based SPAD-array detection. A pulse generator triggers both the pulsed laser and the SPAD-array acquisition, so that each laser pulse is synchronized with one detector frame. Exciton and biexciton photons emitted by a quantum shell are detected in two spatially separated regions of the SPAD array and cross-correlated to obtain $g^{(2)}(0)$. (b) Schematic representation of temporal gating by varying the end time of the integration window. Shortening the integration window reduces the dark-count contribution, but may also reject part of the quantum-shell emission. (c–f) Intensity maps recorded with different integration times: (c) 10 µs, (d) 250 ns, (e) 70 ns, and (f) 20 ns. The comparison illustrates the trade-off between suppressing dark counts and retaining fluorescence signal, with intermediate integration windows preserving most of the emission while strongly reducing the background.

A tunable camera setting determines the duration of the integration window. By choosing this integration time appropriately, the detector can be opened only during the time window in which quantum-shell emission is expected (Figure 3b). The fluorescence lifetime of the quantum shells is on the order of 30 ns (Supporting Information Section S6). If the integration time is much longer (10 µs; Figure 3c), the camera accumulates a significant number of dark counts. If the integration time is too short (Figure 3e,f), the fluorescence is not collected in full. An optimal integration time of 250 ns (Figure 3d) captures the majority of the quantum-shell emission while strongly reducing dark-count coincidences. We therefore use this setting for the high-throughput $g^{(2)}(0)$ measurements.

## High-throughput $g^{(2)}(0)$ mapping

We use the cross-correlation geometry to determine $g^{(2)}(0)$ for many emitting spots in parallel. First, a long-integration reference measurement determines the emitter positions in both detector regions. Emitters are identified with an image-recognition algorithm, and corresponding spots in the two regions are paired based on their relative positions (Supporting Information Section S4). Figure 4a,b show the two detector regions used for cross-correlation, with matched emitter images indicated by circles of the same color.

For each matched spot, we calculate the correlation function $g^{(2)}(\tau)$ with a time resolution equal to the detector integration time (Supporting Information Section S5). Because the SPAD array records whether photons are detected within a predefined integration window, but does not resolve their arrival times within that window, the correlation function is constructed from frame-to-frame photon coincidences. The zero-delay value $g^{(2)}(0)$ is the coincidence count in the same excitation cycle, normalized by the average coincidence count in neighboring excitation cycles. In this way, one $g^{(2)}(0)$ value is obtained for each matched emitting spot.

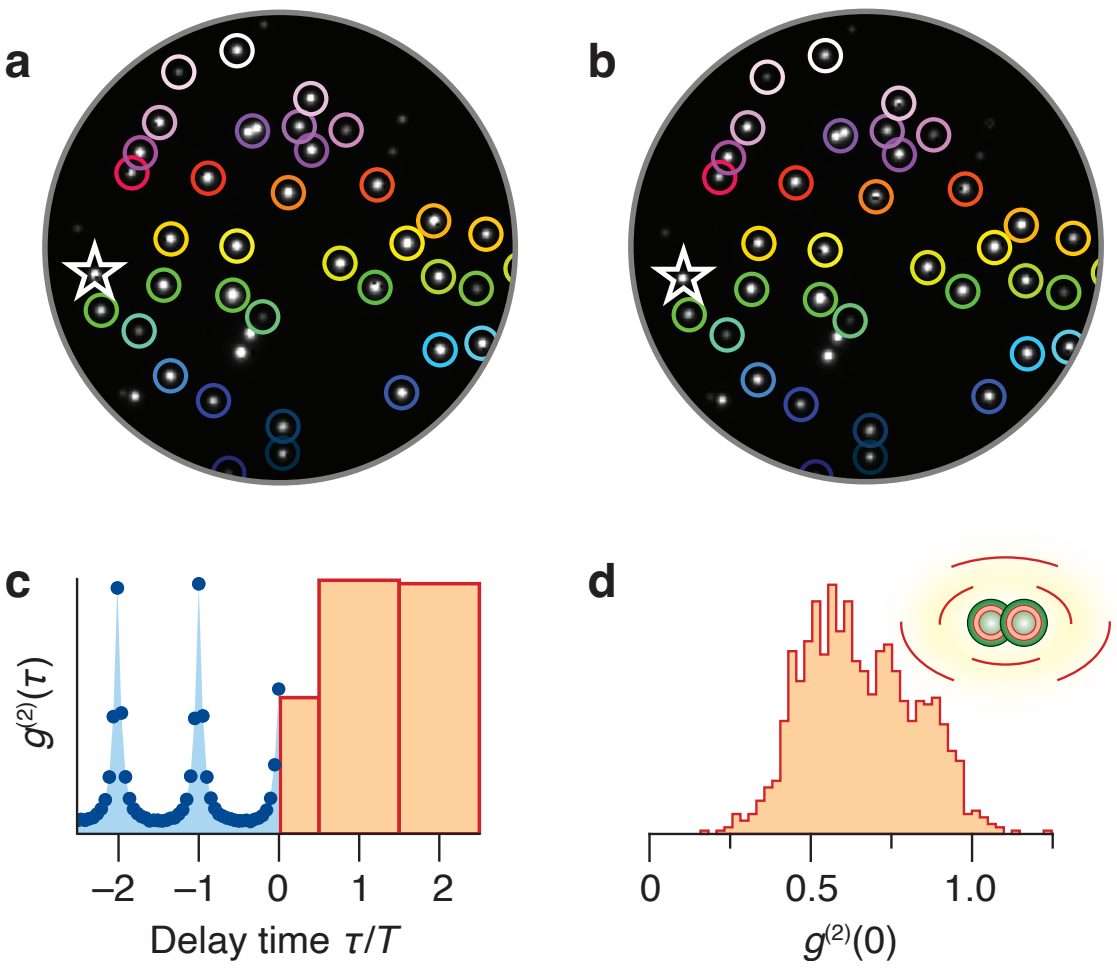

**Figure 4. Measured quantum-shell $\mathbf{g^{(2)}(0)}$ distribution.** (a,b) Intensity maps of the two spatially separated regions on the SPAD array used for cross-correlation. Matched spots assigned to the same quantum shell are indicated by circles of the same color. (c) Comparison of correlation functions measured for the same emitter, marked by the white star in panels a and b, using conventional time tagging (blue) and the frame-based SPAD-array method (red). The extracted values, $g^{(2)}(0) = 0.55$ for time tagging and $g^{(2)}(0) = 0.54$ for the frame-based method, show that frame-based detection preserves the zero-delay-to-side-peak ratio required to quantify biexciton emission. (d) Histogram of $g^{(2)}(0)$ values obtained from more than 1000 matched emitting spots. The low-$g^{(2)}(0)$ population is attributed primarily to individual quantum shells, whereas higher-$g^{(2)}(0)$ contributions are consistent with unresolved clusters.

Figure 4c compares $g^{(2)}(\tau)$ for the same emitter (represented with white star) measured with the frame-based SPAD-array method and with conventional time tagging. In the time-tagged measurement, the pulsed-excitation peaks are temporally resolved, whereas the frame-based method integrates photon coincidences within each 10 µs excitation cycle. The two approaches yield nearly identical biexciton efficiencies, with $\eta_{\mathrm{BX}} = 0.54$ for the frame-based measurement and $\eta_{\mathrm{BX}} = 0.55$ for the time-tagged measurement.

Applying the same frame-based analysis to all matched emitting spots yields the $g^{(2)}(0)$ distribution shown in Figure 4d. This histogram is constructed from 16 measurements on different regions of the substrate, with each measurement consisting of 20 million frames. To exclude poorly sampled particles, only spots with more than 50 coincidences in the side peaks are included. This threshold removes low-intensity spots, for which shot noise leads to a large uncertainty in $g^{(2)}(0)$. The resulting histogram contains 1186 emitting spots and shows a broad, asymmetric distribution with a mean value of 0.65 and a standard deviation of 0.17. The median error on the $g^{(2)}(0)$ values is 0.03, so the width of the distribution reflects actual variations between fluorescence spots to a significant extent.

Several effects can contribute to this broad $g^{(2)}(0)$ distribution. Genuine particle-to-particle variation is expected for chemically synthesized nanocrystals, where differences in size, shell thickness, shape, and composition can alter the balance between radiative recombination and nonradiative Auger recombination.[10–13] In addition, unresolved clusters can increase the measured $g^{(2)}(0)$, because photon pairs may originate from different quantum shells within the same diffraction-limited spot (Supporting Information Section S8.1).[27] The high-$g^{(2)}(0)$ tail may therefore contain cluster-induced coincidences in addition to any intrinsic single-particle heterogeneity.

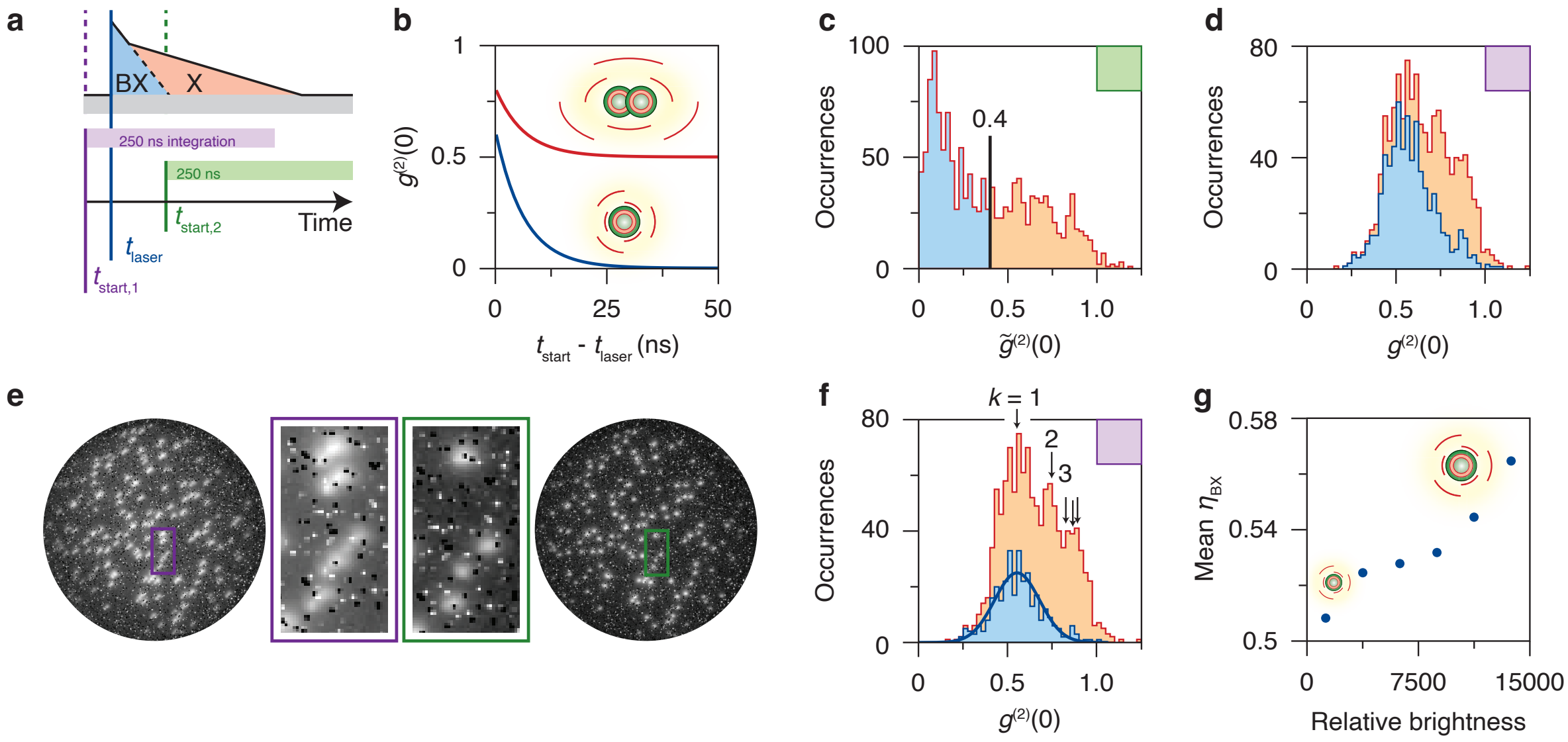

**Figure 5. Identifying single quantum shells by time-gated photon correlation.** (a) Schematic representation of shifting $t_{start}$ in a photon-correlation experiment. Early biexciton emission is suppressed preferentially. (b) Calculated dependence of time-gated $\tilde{g}^{(2)}(0)$ on $t_{start}$ for an isolated quantum shell (blue) and a two-emitter cluster (red). (c) Histogram of $\tilde{g}^{(2)}(0)$, measured for $t_{start}$ of at least 10 ns (varying between pixels). A threshold of $\tilde{g}^{(2)}(0) = 0.4$ separates isolated emitters from clusters. (d) The original $g^{(2)}(0)$ histogram (same as Figure 4d) but now split into individual emitters (blue) and clusters (red) based on the selection in panel c. (e) Intensity maps of the photon-correlation measurement with $t_{start} < t_{laser}$ (left) and of the reference measurement with $t_{start} > t_{laser}$ that identifies clusters. The zoomed regions illustrate possible residual misclassifications due to defocusing, which makes spot overlap, and due to blinking. (f) Blue: selected $g^{(2)}(0)$ distribution for a conservative subset of measurements with minimal drift and defocusing, and with $\tilde{g}^{(2)}(0) < 0.4$. The distribution is an approximately Gaussian single-particle population, with a mean biexciton efficiency of $\eta_{BX} = 0.55$ and an estimated intrinsic standard deviation of 0.12. Arrows: expected peak positions in the $g^{(2)}(0)$ histogram of clusters of $k = 1, 2, 3, 4$, and 5 particles, calculated from the single-particle $\langle \eta_{BX} \rangle = 0.55$ and from the brightness distribution observed (Supporting Information Section S8.4). (g) Mean biexciton efficiency $\langle \eta_{BX} \rangle$ for selected particles, as a function of their fluorescence brightness. Brighter particles have higher $\langle \eta_{BX} \rangle$.

The presence of clusters is indeed evident from the intensity maps in Figure 4a,b, where some emitting spots are non-isolated or partially overlapping. These visible cases represent only the most obvious clusters. Because the optical image is diffraction limited, two quantum shells separated by less than roughly 100 nm can still appear as a single, nearly circular spot. This motivates an additional procedure to distinguish individual quantum shells from clusters.

## Time gating identifies single particles

To recover the intrinsic single-particle biexciton-efficiency distribution, possible cluster contributions must be identified. We use a second form of time gating for this purpose.[27,29] While Figure 3 shortens the integration window by changing $t_{\text{end}}$ to suppress dark counts, here we shift the start of the integration window, $t_{\text{start}}$. This shift suppresses fast biexciton photons more strongly than slower exciton photons (Figure 5a). Consequently, the biexciton–exciton cascades in the zero-delay correlation peak are reduced relative to the correlation side peak. In contrast, coincidences due to photons from different quantum shells in a cluster are suppressed less strongly, because they are dominated by exciton emission. The symbol $\tilde{g}^{(2)}(0)$ denotes the normalized zero-delay correlation for a time-gated experiment where $t_{\text{start}}$ is after the laser pulse $t_{\text{laser}}$.

Figure 5b shows the calculated $\tilde{g}^{(2)}(0)$ for a single quantum shell and for a two-emitter cluster on $t_{\text{start}}$ (Supporting Information Section S8.2 and S8.3). For an individual emitter, increasing $t_{\text{start}}$ removes the biexciton contribution to the zero-delay peak, so $\tilde{g}^{(2)}(0)$ decreases toward zero. For a two-emitter cluster, zero-delay coincidences can still arise from independent exciton photons emitted by different particles, so $\tilde{g}^{(2)}(0)$ remains finite. $\tilde{g}^{(2)}(0)$ approaches 0.5 if the two emitters in the cluster have equal brightness, or slightly lower if the emitters differ in brightness (Supporting Information Section S8.3). A measurement of $\tilde{g}^{(2)}(0)$ can thus be used to distinguish likely isolated emitters from clusters of emitters.

Figure 5c shows the histogram of $\tilde{g}^{(2)}(0)$ for $t_{\text{start}} > 10$ ns (with pixel-to-pixel variations due to hardware limitations; Supporting Information Section S7). These $\tilde{g}^{(2)}(0)$ measurements are recorded separately from the $g^{(2)}(0)$ measurements of Figure 4d but on the same regions of the sample. We use $\tilde{g}^{(2)}(0) < 0.4$ as a conservative criterion (Supporting Information Section S8.3) for selecting individual emitters; spots with $\tilde{g}^{(2)}(0) > 0.4$ are treated as cluster-like.

Figure 5d shows the same data as Figure 4d (red), but now also shows the selected individual emitters (blue) based on this criterion. The $g^{(2)}(0)$ distribution for selected individual emitters is narrower and high-$g^{(2)}(0)$ contributions are reduced.

The $g^{(2)}(0)$ distribution of selected individual emitters still shows a small residual feature near $g^{(2)}(0) \approx 0.8$. While a non-Gaussian biexciton-efficiency distribution is in principle possible, these residual features may also arise from small imperfections in the selection procedure. The measurements with different $t_{\text{start}}$ are acquired separately, and each measurement takes at least 15 min because of camera readout, data transfer, and writing between acquisitions. During this time, emitter blinking could alter the relative intensities of nearby particles and changes in focus can alter the overlap of nearby fluorescence spots (Figure 5e zoom in). This complicates the identification of spots containing one or multiple emitters. Indeed, if we repeat the analysis on a more conservative subset of measurements with minimal drift, the resulting $g^{(2)}(0)$ histogram (Figure 5f) becomes even cleaner. The high-$g^{(2)}(0)$ feature is further reduced and the selected distribution resembles a Gaussian even more. From the $g^{(2)}(0)$ distribution of these selected single quantum shells, we obtain a mean biexciton efficiency of $\langle \eta_{\text{BX}} \rangle = 0.55$ with a standard deviation of $\sigma = 0.13$. The median error is $\sigma_{\text{err}} = 0.04$. From this we estimate the heterogeneity of $\eta_{\text{BX}}$ between quantum shells at $\sigma_{\text{het}} = \sqrt{\sigma^2 - \sigma_{\text{err}}^2} = 0.12$.

Retroactively, we can now confirm our interpretation of high-$g^{(2)}(0)$ spots as clusters.

Using the fitted value of $\langle \eta_{\mathrm{BX}} \rangle$, and accounting for brightness difference between particles (Supporting Information Section S8.4), we calculate where cluster contributions should appear in the $g^{(2)}(0)$ histogram. The predicted cluster positions (arrows in Figure 5f) match features in the histogram, without introducing additional fit parameters.

We hypothesized that the variation of $\eta_{\mathrm{BX}}$ may reflect variations in overall particle size. Indeed, the Auger recombination rate is known to follow volume scaling: larger particles have weaker Auger recombination and therefore more efficienct multi-exciton emission.[30,32] Checking size correlations directly would require optical measurements and electron microscopy on each individual particle. Here, we use a simpler approximate method: we treat the fluorescence brightness as an optically accessible proxy for particle size. A larger particle absorbs more, so it is on average brighter.[33] To estimate the intrinsic brightness of a particle, we normalize its fluorescence count rate to the local fluorescence background count rate. This normalization divides out the influence of the local excitation laser intensity.

The extracted $\langle \eta_{\mathrm{BX}} \rangle$ as a function of particle brightness (Figure 5g) shows a clear correlation. Inspecting the histograms from which these mean values are fitted confirms that the trend is due to a shift of the single-particle distribution rather than due to increasing cluster contributions (Supporting Information Section S9). The trend is qualitatively consistent with the volume scaling of Auger recombination. Quantitative analysis would require future work, considering blinking and dipole orientation more carefully, which distorts the simple brightness metric that we used here.[31,34]

# Conclusions

This work presents a strategy for artifact-free high-throughput photon-correlation imaging of quantum emitters. SPAD-array photon correlation quantifies biexciton-efficiency heterogeneity across more than 1000 quantum-shell spots. Spatially separating two images of the same sample is key to suppressing artifacts due to short-range crosstalk between pixels, while time gating reduces dark-count coincidences and enables identification of single particles and clusters. The full analysis procedure yields a single-particle biexciton-efficiency distribution with a mean of 0.55 and an estimated intrinsic standard deviation of 0.12. The raw $g^{(2)}(0)$ histogram is consistent with various cluster sizes of a finite number of particles. Finally, we find a correlation of biexciton efficiency with particle brightness, which is consistent with the volume scaling of Auger recombination. Future method development could aim for the acquisition of spectral information[35] alongside photon correlations, to either unravel the energetics of multi-excitons or as an additional proxy for structural properties of particles.

# Experimental Methods

## High-throughput second-order correlation measurements

All measurements were performed on a custom-built microscope based on a Nikon Ti-U inverted microscope. Single quantum-shell samples were prepared by spin-coating a highly diluted quantum-shell solution onto a glass coverslip. The dilution was approximately 1 : 20000 relative to the stock solution. The coverslip was mounted onto a microscope slide using an airtight adhesive spacer and placed on the microscope.

The sample was excited in widefield geometry using a pulsed 405 nm laser diode (PicoQuant LDH-D-C-405, driven by a PicoQuant PDL 800-D controller). The excitation beam was directed to the microscope through a dichroic mirror (425 nm, Thorlabs DMLP425R) and collimated by a high-numerical-aperture oil-immersion objective (Nikon CFI Plan Apochromat Lambda 100×, NA 1.45) such that it illuminated a large part of the field of view. Emission was collected through the same objective and separated from the excitation light by the dichroic mirror.

The collected fluorescence was split into two paths using a 50:50 non-polarizing beamsplitter (Thorlabs BS013). The two beams were projected onto spatially separated regions of the SPAD-array detector. In this configuration, each quantum shell appears twice on the detector, and photon correlations can be calculated between the two corresponding images of the same emitter, while the large spatial separation between the two images suppresses short-range detector crosstalk.

A long-integration reference measurement was recorded to determine the positions of the quantum shells in the field of view. This reference measurement was acquired at a laser repetition rate of 10 MHz with a total integration time of 1 min.

For the high-throughput $g^{(2)}(0)$ measurements, the laser and SPAD-array detector (SPAD512, Pi Imaging) were synchronized using a pulse generator (Aim-TTi TGA1244). The pulse generator triggered both the laser and the detector so that each laser pulse corresponded to one detector frame. The detector was operated at its maximum frame rate of 80 kHz, with 1-bit readout per pixel. Each measurement consisted of 100.000 frames and was repeated 200 times using an automated Python acquisition script, resulting in 20 million excitation cycles per field of view.

Two frame-based photon-correlation measurements were recorded for each field of view. In the first measurement, the detector gate was opened without delaying the detector trigger relative to the laser pulse, using an integration time of 250 ns. This measurement was used to determine the main $g^{(2)}(0)$ distribution. In the second measurement, the detector trigger was delayed by 120 ns relative to the laser pulse. This time-gated measurement was used to distinguish likely individual emitters from unresolved clusters of emitters.

For each measurement, photon coincidences were calculated between the two spatially separated images of the same sample on the SPAD array. The zero-delay coincidence count was obtained from detections occurring in corresponding emitter regions within the same detector frame, whereas side-peak coincidences were obtained from detections in different frames. The normalized $g^{(2)}(0)$ value for each emitter was calculated as the ratio of the zero-delay coincidence count to the average side-peak coincidence count.

## Associated content

The Supporting Information contains quantum-shell synthesis procedure, gain calculations, SPAD-array crosstalk and delay calibrations, image processing and emitter matching, frame-based $g^{(2)}(0)$ analysis, dark-count correction, cluster and delayed-gate models, and brightness-dependent biexciton-efficiency analysis.

## Funding

T.A. and F.T.R. were supported by the Dutch Research Council NWO (Vi.Vidi.203.031). D.H., D.N. and M.Z. acknowledge the support by NSF Award CHE-2505151.

Supporting Information for

# Quantifying the Distribution of Biexciton Emission Efficiencies in Colloidal Quantum Shells

Tjom Arens[1], Dulanjan Harankahage[2,3], Divesh Nazar[2,3], Mikhail Zamkov[2,3], and Freddy T. Rabouw[1,*]

[1] Debye Institute for Nanomaterials Science, Utrecht University, Princetonplein 1, 3584CC Utrecht, The Netherlands
[2] The Center for Photochemical Sciences, Bowling Green State University, Bowling Green, Ohio 43403, USA
[3] Department of Physics, Bowling Green State University, Bowling Green, Ohio 43403, USA
*Corresponding author: f.t.rabouw@uu.nl

# S1 Synthesis and characterization

**Materials.** All chemicals were used as received without further purification. The following reagents were used: anhydrous acetone, 99% (Amresco); cadmium oxide, CdO, 99.95% (MilliporeSigma); zinc acetate dihydrate, 98% (Acros Organics); ethanol, 99% (BeanTown Chemical); hexane (Thermo Scientific); 1-octadecene, ODE, 90% (MilliporeSigma); octane, 98% (MilliporeSigma); 1-octanethiol, 97% (Alfa Aesar); oleic acid, OA, 90% (MilliporeSigma); oleylamine, OLAM, 70% (MilliporeSigma); dioctylamine, DOA, 97% (MilliporeSigma); selenium powder, Se, 99.5%, 200 mesh (Thermo Scientific); sulfur powder, S, 99.999% (Thermo Scientific); toluene, 99.8% (MilliporeSigma); and tri-n-octylphosphine, TOP, 97% (Strem Chemical).

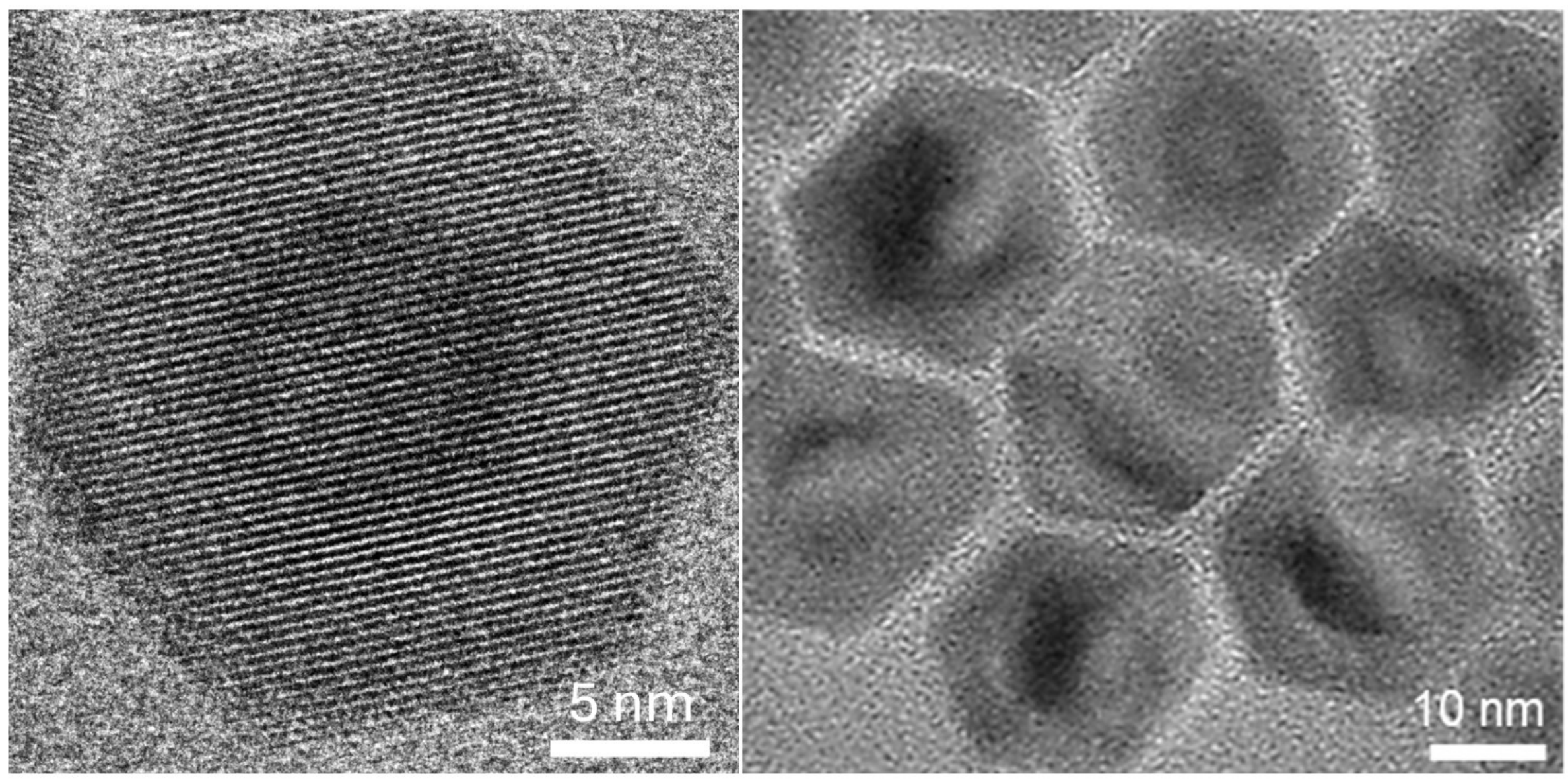

**Figure S1. TEM images of CdS/CdSe/CdS quantum shells.** The particle diameters during growth were $7.9 \pm 0.8$ nm for the CdS cores, $11.9 \pm 0.8$ nm after CdSe shell growth, and $21.5 \pm 1.4$ nm after growth of the outer CdS shell.

**Synthesis of 6–9 nm Bulk-Sized CdS nanocrystals.** Large CdS nanocrystals were synthesized using a coalescence-based growth procedure.[1] In a typical reaction, 8 mL of OLAM and 42 mg of CdCl2 were loaded into a 25 mL three-neck flask, transferred to an argon atmosphere using a Schlenk line, and heated to 240 °C. Once the target temperature was reached, 540 nmol of small CdS nanocrystals, 2–4 nm in diameter, was rapidly injected into the flask. The reaction was allowed to proceed for approximately 60 min, during which rapid coalescence growth occurred. After the growth period, the flask was removed from the heating mantle and allowed to cool. The resulting nanocrystals were purified by precipitation using a toluene/ethanol mixture. First, an equal volume of ethanol was added to the crude nanocrystal solution in a centrifuge tube. The tube was gently inverted twice to mix the solution, avoiding vigorous shaking. The mixture was then centrifuged for 2–3 min at 4500 rpm. The supernatant was carefully discarded without disturbing the nanocrystal pellet. The pellet was redispersed in fresh hexane, and the washing process was repeated. After the final wash, the supernatant was removed completely, and the nanocrystal pellet was allowed to air-dry. The dried product was then redispersed in hexane for further use. The CdS core nanocrystals were subsequently recapped with OA ligands. For this step, the purified CdS nanocrystals were dissolved in a mixture of 5 mL of ODE and 8 mL of OA. The flask was placed under argon and heated to 150 °C for 60 min. The final product was purified, redispersed in hexane, and stored under ambient conditions.

**Synthesis of CdS/CdSe Core/Shell nanocrystals.** Growth of the CdSe shell was carried out by co-injection of Cd and Se precursors using two syringe pumps. A 0.1 M Cd-oleate precursor was prepared by combining 412 mg of CdO, 8 mL of OA, and 5 mL of ODE in a 50 mL flask. The mixture was heated to 260 °C under argon until a clear, nearly colorless solution was obtained. At this point, an additional 19 mL of ODE was added to the flask. The 0.1 M TOP-Se precursor was prepared by combining 141 mg of Se powder and 3 mL of TOP in a 25 mL flask. The mixture was heated to 140 °C under argon until the selenium powder had fully dissolved. Subsequently, 14 mL of ODE was added to dilute the precursor. To initiate CdSe shell growth, 540 or 1080 nmol of CdS nanocrystals was loaded into a 100 mL flask containing 2 mL of DOA and 2 mL of ODE. The mixture was heated to 110 °C and degassed until bubbling ceased. The flask was then placed under argon and heated to 315 °C. When the temperature reached 270 °C, injection of the Cd-oleate and TOP-Se precursors was initiated at a rate of 3 mL $h^{1}$. The injection was continued until the photoluminescence feature reached the desired wavelength, typically between 620 and 695 nm. The total injection time was approximately 90 min for larger CdS nanocrystals and approximately 70 min for medium-sized CdS nanocrystals. After the injection was complete, the reaction flask was removed from the heating mantle and cooled to room temperature. The nanocrystal product was purified using the precipitation procedure described above.

**Synthesis of CdS/CdSe/CdS Core/Shell/Shell Nanocrystals.** The Cd-oleate precursor was prepared as described above. The sulfur precursor was prepared by mixing 11.66 mL of ODE with 0.34 mL of 1-octanethiol. Separately, 540 or 1080 nmol of previously synthesized CdS/CdSe core/shell nanocrystals was loaded into a 100 mL flask containing 2 mL of DOA and 2 mL of ODE. The mixture was degassed at 130 °C, transferred to an argon atmosphere, and heated to 315 °C. The Cd-oleate and sulfur precursors were injected using separate syringe pumps at a rate of 3 mL $h^{1}$, beginning when the reaction temperature reached 270 °C. The shell-growth reaction was typically continued for 2–4 h. After completion of the injection, the reaction mixture was annealed for an additional 45 min and then cooled to room temperature. The resulting CdS/CdSe/CdS nanocrystals were purified using the procedure described above and redispersed in hexane.

**Deposition of a ZnS Shell through CdS/ZnS Alloying.** The ZnS shell was deposited by sequential injection of three precursors: 0.1 M TOP-S, 0.1 M Cd-oleate, and 0.1 M Zn-oleate by adapting a previous protocol[2,3] . The Cd-oleate precursor was prepared as described above and maintained under argon at 100 °C. The TOP-S precursor was prepared by combining 38.4 mg of sulfur powder and 2 mL of TOP in a 25 mL flask. The mixture was heated to 140 °C under argon until the sulfur had fully reacted, after which 10 mL of ODE was added to dilute the solution. The Zn-oleate precursor was prepared by combining 526 mg of zinc acetate dihydrate, 3 mL of OA, and 15 mL of ODE in a 50 mL flask. The mixture was degassed stepwise at 60, 80, 100, and 120 °C. After degassing, the flask was placed under argon, and 2 mL of TOP was injected. The temperature was then increased to 240 °C, after which the solution was cooled and maintained at 100 °C. For shell growth, 540 or 1080 nmol of CdS/CdSe/CdS nanocrystals were loaded into a 100 mL flask containing 2 mL of ODE and 2 mL of DOA. The mixture was degassed at 110 °C, placed under argon, and heated to 260 °C. During heating, a 3:1 mL mixture of Cd-oleate and Zn-oleate precursors was prepared in a 10 mL flask under argon. This Cd/Zn precursor mixture was loaded into a syringe and paired with an equal volume of the TOP-S precursor in a second syringe. The two precursors were injected into the reaction flask for 30 min at a rate of 3 mL h1. After the first injection, the Cd/Zn syringe was removed, flushed with hexane, and reloaded with a 2:2 mL mixture of Cd-oleate and Zn-oleate. The sulfur syringe was reloaded with a matching volume of TOP-S, and the injection was continued for another 30 min at 3 mL h1. This process was then repeated using a 1:3 mL mixture of Cd-oleate and Zn-oleate. Finally, a 100% Zn-oleate precursor and TOP-S were injected for 40 min. Immediately after the final injection began, the reaction temperature was increased to 315 °C. Once the injection was complete, 1.5 mL of OA was added to the flask, and the heating mantle was turned off to allow slow cooling. The final nanocrystals were precipitated with toluene and a 1:2 ethanol/acetone mixture by centrifugation and then redispersed in hexane.

# S2 Gain calculations for homogeneous and heterogeneous biexciton-efficiency distributions

A simple state-filling model illustrates the effect of variations in biexciton emission efficiency on the gain of quantum dots (or quantum shells). In the example discussed in the main text, we compare two limiting ensembles with the same average biexciton efficiency, $\langle \eta_{\mathrm{BX}} \rangle$, but with different distributions. In the homogeneous ensemble, every emitter has the same biexciton efficiency. In the heterogeneous ensemble, a fraction of the emitters have unity biexciton efficiency, $\eta_{\mathrm{BX}} = 1$, while the other emitters have fully quenched biexciton emission, $\eta_{\mathrm{BX}} = 0$.

We modeled optical gain using band-edge state filling.[4] We took the band-edge electron and hole degeneracies as $g_{\mathrm{e}} = 2$ and $g_{\mathrm{h}} = 4$, respectively. These degeneracies represent the number of available band-edge electron and hole states that can participate in absorption and stimulated emission. For an $n$-exciton state, we define the number of occupied electron states as $n_{\mathrm{e}}(n)$, and the number of occupied hole states as $n_{\mathrm{h}}(n)$. These occupancies increase with exciton number until the corresponding band-edge degeneracy is filled:

$$n_{\mathrm{e}}(n) = \begin{cases} n & ; \quad n < g_{\mathrm{e}} \\ g_{\mathrm{e}} & ; \quad n \geq g_{\mathrm{e}} \end{cases} \tag{S1}$$

and

$$n_{\mathrm{h}}(n) = \begin{cases} n & ; \quad n < g_{\mathrm{h}} \\ g_{\mathrm{h}} & ; \quad n \geq g_{\mathrm{h}} \end{cases} \tag{S2}$$

The number of unoccupied electron and hole states is $g_{\mathrm{e}} - n_{\mathrm{e}}(n)$ and $g_{\mathrm{h}} - n_{\mathrm{h}}(n)$, respectively. We wrote the net absorption contribution of the $n$-exciton state as

$$A_n = [g_{\mathrm{e}} - n_{\mathrm{e}}(n)]\,[g_{\mathrm{h}} - n_{\mathrm{h}}(n)] - n_{\mathrm{e}}(n) n_{\mathrm{h}}(n). \tag{S3}$$

The first term counts the number of available band-edge absorption pathways, corresponding to transitions into empty electron and hole states. The second term counts the number of occupied electron–hole recombination pathways that contribute to stimulated emission. Positive values of $A_n$ correspond to net absorption, whereas negative values correspond to net gain.

## Steady-state gain for continuous-wave excitation

For continuous-wave excitation, we calculate the steady-state occupation probabilities from a balance between optical absorption and recombination. We assume that the radiative recombination rate of the $n$-exciton state scales with the number of occupied electron–hole recombination pathways,

$$k_{\mathrm{r},n} = n_{\mathrm{e}}(n) n_{\mathrm{h}}(n). \tag{S4}$$

The biexciton efficiency $\eta_{\mathrm{BX}}$ depends on the balance between radiative decay (rate constant $k_{\mathrm{r},2}$) and Auger decay (rate constant $k_{\mathrm{A},2}$) of the biexciton state:[5]

$$\eta_{\mathrm{BX}} = \frac{k_{\mathrm{r},2}}{k_{\mathrm{r},2} + k_{\mathrm{A},2}}. \tag{S5}$$

We use this relation to express the Auger decay rate in terms of the biexciton efficiency:

$$\frac{k_{\mathrm{A},2}}{k_{\mathrm{r},2}} = \frac{1-\eta_{\mathrm{BX}}}{\eta_{\mathrm{BX}}}. \tag{S6}$$

For higher multiexciton states, the Auger decay rates increase with exciton number according to

$$k_{\mathrm{A},n} = n^2(n-1)\frac{1-\eta_{\mathrm{BX}}}{\eta_{\mathrm{BX}}}. \tag{S7}$$

This scaling makes the Auger rate increase more rapidly with exciton number than the radiative decay rate.[6] Physically, this reflects that Auger recombination involves at least three carriers: one electron–hole pair recombines nonradiatively and transfers its energy to an additional electron or hole. Radiative recombination, in contrast, involves the recombination of only one electron–hole pair. Consequently, higher multiexciton states have a lower emission efficiency than the biexciton state, because nonradiative Auger recombination becomes increasingly dominant as the number of carriers increases.

We calculate the steady-state population of the $n$-exciton state recursively as

$$p_n = B\frac{k_{\mathrm{abs}}^n}{\prod_{m=1}^{n}\left(k_{\mathrm{r},m} + k_{\mathrm{A},m}\right)} \tag{S8}$$

with normalization constant $B$ such that $\sum_{n=0}^{\infty} p_n = 1$. The absorption rate, $k_{\mathrm{abs}}$, does not depend on $n$ because we consider non-resonant absorption where bleach effects are negligible. This expression describes the competition between optical pumping, which drives the system into higher exciton states, and recombination, which returns the system to lower exciton states. Strong Auger recombination suppresses the population of multiexciton states and therefore reduces the ability of the ensemble to sustain optical gain.

We then calculate the continuous-wave gain signal as

$$G_{\mathrm{CW}} = \sum_n p_n A_n. \tag{S9}$$

This signal is positive when absorption dominates and negative when stimulated emission dominates. Figure 1c in the main text compares the quantity $G_{\mathrm{CW}}$ as a function of $k_{\mathrm{abs}}$, for emitter ensembles with heterogeneous and homogeneous biexciton efficiency with $\langle\eta_{\mathrm{BX}}\rangle = 0.5$.

## Gain lifetime for pulsed excitation

For pulsed excitation, we assume that the number of absorbed photons per nanocrystal follows Poisson statistics immediately after the excitation pulse, at $t = 0$:

$$p_n(0) = \exp(-\mu)\frac{\mu^n}{n!}, \tag{S10}$$

where $\mu$ is the average number of absorbed photons per pulse. This gives the initial population of each $n$-exciton state directly after excitation. The calculation shown in the main text uses $\mu = 1.7$. Following the pulse, multi-exciton populations evolve through recombination. We describe this decay with a rate matrix $\mathbf{\Gamma}$, which transfers population sequentially from the $n$-exciton state to the $(n-1)$-exciton state. The total decay rate of each $n$-exciton state is

$$k_{\mathrm{tot},n} = k_{\mathrm{r},n} + k_{\mathrm{A},n}. \tag{S11}$$

The diagonal elements of $\mathbf{\Gamma}$ describe loss of population from each $n$-exciton state,

$$\Gamma_{n,n} = -k_{\mathrm{tot},n}, \tag{S12}$$

whereas the off-diagonal elements describe population arriving in the next lower exciton state after recombination,

$$\Gamma_{n-1,n} = k_{\mathrm{tot},n}. \tag{S13}$$

All other elements are zero. We calculate the time-dependent population vector as

$$\mathbf{p}(t) = \exp(\mathbf{\Gamma}t)\mathbf{p}(0), \tag{S14}$$

where $\mathbf{p}(0)$ contains the Poisson-distributed initial exciton populations. This matrix exponential is the solution of the coupled rate equations

$$\frac{\mathrm{d}\mathbf{p}(t)}{\mathrm{d}t} = \mathbf{\Gamma}\mathbf{p}(t). \tag{S15}$$

This step calculates how the initially excited ensemble relaxes in time as multiexciton states decay into lower exciton states. Particles with low $\eta_{\mathrm{BX}}$ have large Auger rates and therefore rapidly lose their multiexciton population. Particles with high $\eta_{\mathrm{BX}}$ have suppressed Auger recombination and therefore retain gain-producing multiexciton states for longer.

At each time point, we calculate the gain signal from the state populations,

$$G(t) = \sum_n A_n p_n(t). \tag{S16}$$

Here, $A_n$is the net absorption contribution of the $n$-exciton state. Negative values of $G(t)$ correspond to net gain, whereas positive values correspond to net absorption. We define the gain lifetime $\tau_{\mathrm{gain}}$ as the first time after the pulse at which $G(t)$ crosses zero from negative to positive. Figure 1d of the main text plots $\tau_{\mathrm{gain}}$ as a function of $\langle\eta_{\mathrm{BX}}\rangle$.

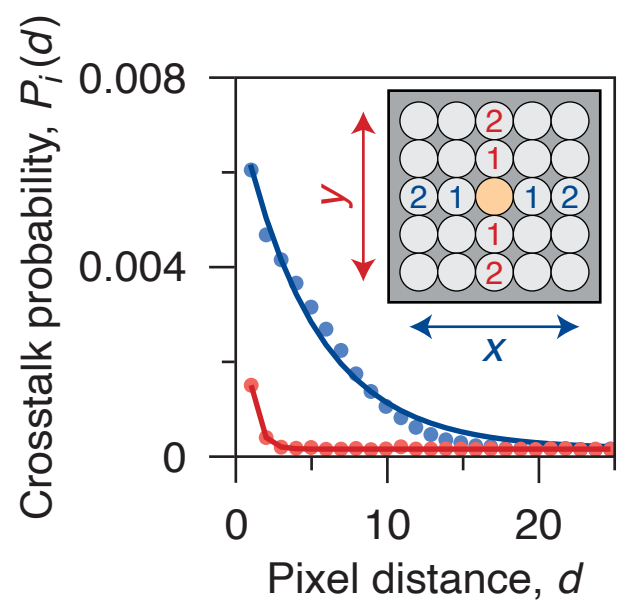

**Figure S2. Spatial dependence of SPAD-array crosstalk.** Crosstalk probability as a function of axial pixel separation for the two detector directions. The probabilities were obtained from measurements with an uncorrelated light source by calculating the conditional probability that a second pixel registers a photon in the same frame at a distance $d$ from a detected event. The horizontal and vertical directions were analyzed separately, yielding $P_x(d)$ and $P_y(d)$, respectively. Solid lines show exponential fits with a fixed background level determined from the average of the final five distance points.

# S3 Crosstalk calculations

To quantify the crosstalk probability as a function of pixel separation, we illuminated the detector with an uncorrelated light source.[7,8] For each detected photon event, we evaluated whether another pixel at an axial distance $d$ registered a photon in the same frame. This gives the conditional probability that a detection event is accompanied by a second detection event at a defined pixel separation.

We analyzed the two detector directions separately (Figure S2). We calculated the horizontal coincidence probability, $P_x(d)$, from events separated along the $x$-direction, and the vertical coincidence probability, $P_y(d)$, from events separated along the $y$-direction. The two directions show a clear asymmetry: the nearest-neighbor crosstalk probability in the $x$-direction is approximately four times higher than in the $y$-direction.

The measured probabilities contain a short-range crosstalk contribution and a distance-independent accidental-coincidence background from the uncorrelated light source. We therefore fitted the distance dependence with an exponential decay plus a constant background,

$$P_i(d) = A_i \exp\left(-\frac{d}{\lambda_i}\right) + B_i, \tag{S17}$$

where $i = x, y$. Here, $A_i$ is the short-range crosstalk amplitude, $\lambda_i$ is the decay length in pixels, and $B_i$ is the accidental-coincidence background. We fixed $B_i$ to the average probability of the final five distance points for each direction and fitted the remaining short-range component by varying $A_i$ and $\lambda_i$.

The fitted parameters were

$$A_x = 7.28 \times 10^{-3}, \qquad \lambda_x = 5.02 \text{ pixels}, \tag{S18}$$

and

$$A_y = 7.46 \times 10^{-3}, \qquad \lambda_y = 0.59 \text{ pixels}. \tag{S19}$$

Although the fitted amplitudes are similar, the longer decay length in the $x$-direction causes the crosstalk probability to remain higher over several pixels. In both directions, however, the short-range crosstalk decays over distances far below the separation used in the cross-correlation experiment. In the experiments of the main text (Figs. 2–5), the corresponding images of the same quantum shell are separated by more than 200 pixels on the SPAD array. This separation is far outside the measured crosstalk range, showing that detector crosstalk cannot generate coincidences between the two spatially separated images used to calculate $g^{(2)}(0)$.

# S4 Image processing, emitter localization, and spot matching

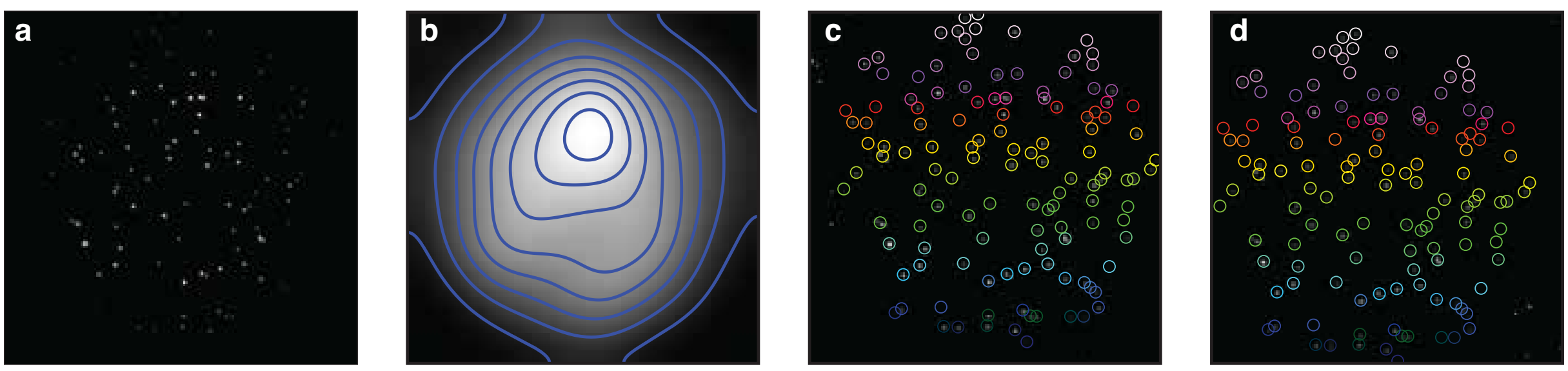

**Figure S3. Emitter localization and matching of spatially separated images.** (a) Long-integration reference image of quantum shells on a glass slide at low density, before background correction and normalization. (b) Estimated background used for background correction and relative-brightness normalization. (c,d) Background-corrected intensity maps of the two spatially separated detector regions. Matched images of the same emitter are indicated by circles of the same color.

We determined emitter positions from long-integration reference measurements in which the same sample was imaged onto two spatially separated detector regions (Figure S3a). For each image, we estimated the local background by convolving the image with a Gaussian kernel, which replaces each pixel by a distance-weighted average of the surrounding pixels (Gaussian smoothing; $\sigma = 10$ pixels). Bright particle positions were identified as pixels more than $3\sigma$ above this local background estimate and were removed from the background image. The removed pixels were filled by interpolation from the surrounding pixels using Gaussian-weighted averaging (Gaussian inpainting). We repeated this smoothing, masking, and interpolation procedure 10 times to obtain the background estimation (Figure S3b). We then subtracted the estimated background from each raw image and normalized the result by the same background image to obtain the corrected images (Figure S3c,d).

Next, we detected emitting spots in each corrected image by thresholding pixels more than one standard deviation above the image mean and identifying connected components above this threshold. For each object, we calculated the intensity-weighted centroid and a bounding box. We used the centroid to match corresponding spots between the two detector regions, while the bounding box defined the pixels used for photon counting in the correlation analysis.

Correctly matching the two images of the same emitter is essential, because we calculate the photon-correlation function from the cross-correlation between these two spatially separated images. If spot $i$ in the first detector region were matched to the wrong spot $j$ in the second detector region, the resulting correlation would combine photons from different quantum shells and would therefore not represent the $g^{(2)}(0)$ of a single emitter.

We matched emitter spots between the two detector images by finding the globally optimal one-to-one assignment. Let $\mathbf{r}_i$ be the centroid position of emitter $i$ in the first detector region, and let $\mathbf{s}_j$ be the centroid position of emitter $j$ in the second detector region. Because the two detector regions image the same sample, the two emitter patterns are related approximately by a translation $\mathbf{t} = (\Delta r, \Delta c)$, where $\Delta r$ and $\Delta c$ are the row and column displacements. We first estimated this translation from a histogram of all pairwise offsets,

$$\mathbf{D}_{ij} = \mathbf{r}_i - \mathbf{s}_j \quad \text{for all } (i, j), \tag{S20}$$

with a bin width of 1 pixel. The most frequently occurring offset gives an initial estimate $\mathbf{t}_0$ of the displacement between the two detector regions.

After this initial translation estimate, we shifted the centroids in the second detector region into the coordinate system of the first region,

$$\mathbf{s}'_j = \mathbf{s}_j + \mathbf{t}_0. \tag{S21}$$

We then updated the displacements between each pair of spots and calculated the updated scalar distance

$$D'_{ij} = \left|\mathbf{r}_i - \mathbf{s}'_j\right| \quad \text{for all } (i, j). \tag{S22}$$

We used this distance matrix as the cost matrix for the Hungarian assignment algorithm.[9] The algorithm finds the one-to-one assignment $M^*$ that minimizes the total matching cost,

$$M^* = \arg\min_{M} \sum_{(i,j)\in M} D'_{ij}, \tag{S23}$$

with the constraint that each spot in each detector region can be used at most once. We excluded candidate pairs with $D'_{ij}$ larger than 10 pixels. This avoids assigning one spot to multiple partners and gives a globally optimal set of matched emitter pairs rather than a sequence of independent nearest-neighbor decisions.

We refined the translation iteratively. After each assignment step, we recalculated the displacement between the emitter pairs

$$\mathbf{D}_{ij} = \mathbf{r}_i - \mathbf{s}_j \quad \text{for } (i, j) \in M^*. \tag{S24}$$

We then updated the translation estimate using the component-wise median displacement of the matched emitter pairs,

$$\mathbf{t}_{\text{new}} = \Big(\text{median}_{(i,j)\in M^*} \left[r_i - s_j\right]_r, \ \text{median}_{(i,j)\in M^*} \left[r_i - s_j\right]_c\Big), \tag{S25}$$

We used the median to reduce the influence of occasional incorrect matches. We then recalculated the shifted coordinates, distance matrix, and Hungarian assignment using the updated translation. We repeated this procedure five times, yielding a stable set of matched emitter pairs (Figure S3c,d).

For each matched emitter pair, we defined two rectangular bounding boxes, one in each detector region, using the smallest box that contained all connected pixels of the corresponding spot. Before calculating the correlation functions, we corrected for small residual shifts between the long-integration reference image and the intensity map of the frame-based correlation measurement. We generated an intensity map from the correlation measurement by summing photon detections over all frames. The bounding boxes obtained from the reference image were then shifted by integer row and column offsets, and for each offset we calculated the total correlation-measurement intensity contained within all bounding boxes. The offset that maximized this summed intensity was used to align the bounding boxes to the correlation measurement. This final shift correction ensured that the pixels used for photon counting were centered on the emitter positions during the actual $g^{(2)}(0)$ acquisition.

For each frame, we then determined whether a photon was detected within the aligned bounding boxes for each of the two spatially separated images. This converted the SPAD data into two binary photon-detection traces for each emitter, which were used to calculate the frame-based cross-correlation.

# S5 Frame-based $g^{(2)}(0)$ calculation

For each matched emitter, we used the photon-count traces from the two spatially separated images to calculate a frame-based cross-correlation function. Let $A_p(t)$ and $B_p(t)$ denote the binary (i.e., valued 0 or 1) photon-detection traces for particle $p$ in the two detector regions. We calculated the coincidence count at lag $l$ as

$$C_p(l) = \sum_t A_p(t+l)B_p(t). \tag{S26}$$

Here, both $t$ and $l$ are discrete times, at integers of the repetition period of the laser. The zero-delay coincidence count is $C_p(0)$. We obtained side-peak counts from coincidences between neighboring excitation cycles, $C_p(l > 0)$. We calculated the frame-based zero-delay correlation value as

$$g_p^{(2)}(0) = \frac{C_p(0)}{\langle C_p(l)\rangle_{l\in\{1,2,3,4,5\}}}, \tag{S27}$$

where $\langle C_p(l)\rangle_{l\in\{1,2,3,4,5\}}$ is the mean of the side-peak coincidence counts with $1 \le l \le 5$.

# S6 Fluorescence lifetime analysis and dark-count correction

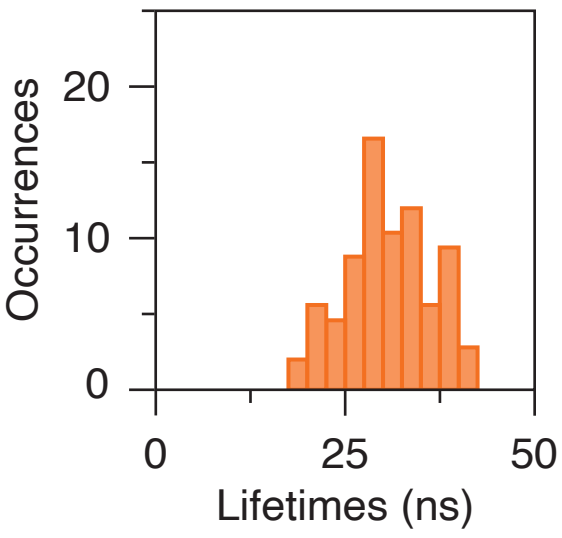

**Figure S4. Fluorescence lifetime distribution of quantum shells.** Fluorescence lifetimes were obtained from gated decay measurements by fitting the aligned decay traces with a monoexponential function. The distribution is centered around 30 ns, supporting the use of a 250 ns integration window for the frame-based $g^{(2)}(0)$ measurements.

We used temporal gating to restrict the detector integration window to the time range in which quantum-shell emission is expected. To determine the relevant emission timescale, we recorded gated movies in which the detector gate was shifted through the fluorescence decay. For each gate position, we summed the signal within the bounding boxes corresponding to the two spatially separated images of each emitter.

Small timing offsets can occur between pixels or detector regions. We therefore aligned the decay curves in time by cross-correlation to a reference pixel before summing them. The aligned decay traces were fitted with a monoexponential function,

$$I(t) = A \exp\left(-\frac{t}{\tau}\right) + C, \tag{S28}$$

where $A$ is the decay amplitude, $\tau$ is the fluorescence lifetime, and $C$ is a constant background. This procedure gave a fluorescence lifetime for each particle, with a distribution centered at around 30 ns (Figure S4).

We used these lifetime estimates to choose the 250 ns integration window for the main frame-based $g^{(2)}(0)$ measurements. This window captures more than 99% of the quantum-shell emission, while reducing the active integration time by a factor of 50 compared with the full 12.5 $\mu$s excitation period. Because the dark-count contribution scales with the active integration time, this temporal gate strongly suppresses dark-count coincidences while retaining nearly all fluorescence photons.

Although the 250 ns gate strongly reduces dark counts, a residual contribution remains. We estimated this residual contribution from dark measurements and converted it into a dark-count probability per frame for each particle bounding box in both detector regions. For a given lag $l$, the observed coincidence count contains true signal–signal coincidences, but also signal–dark, dark–signal, and dark–dark coincidences. We therefore corrected the observed coincidence count as

$$C_{\mathrm{sig}}(l) = C_{\mathrm{obs}}(l) - p_{\mathrm{d},B} \sum_t A(t+l) - p_{\mathrm{d},A} \sum_t B(t) + N_l p_{\mathrm{d},A} p_{\mathrm{d},B}. \tag{S29}$$

Here, $C_{\mathrm{obs}}(l)$ is the observed coincidence count at lag $l$, $C_{\mathrm{sig}}(l)$ is the dark-corrected signal coincidence count, and $A(t)$ and $B(t)$ are the binary photon-detection traces from the two detector regions. The dark-count probabilities per frame in the two corresponding bounding boxes are $p_{\mathrm{d},A}$ and $p_{\mathrm{d},B}$, and $N_l$ is the number of valid frame pairs contributing to $l$.

The second and third terms remove coincidences in which one detection is caused by quantum-shell emission and the other by a dark count. The final term adds back the dark-dark contribution, which is subtracted twice by the preceding two terms. We applied this correction separately for each particle and each correlation lag before calculating $g^{(2)}(0)$.[7,10]

# S7 SPAD delay calibration

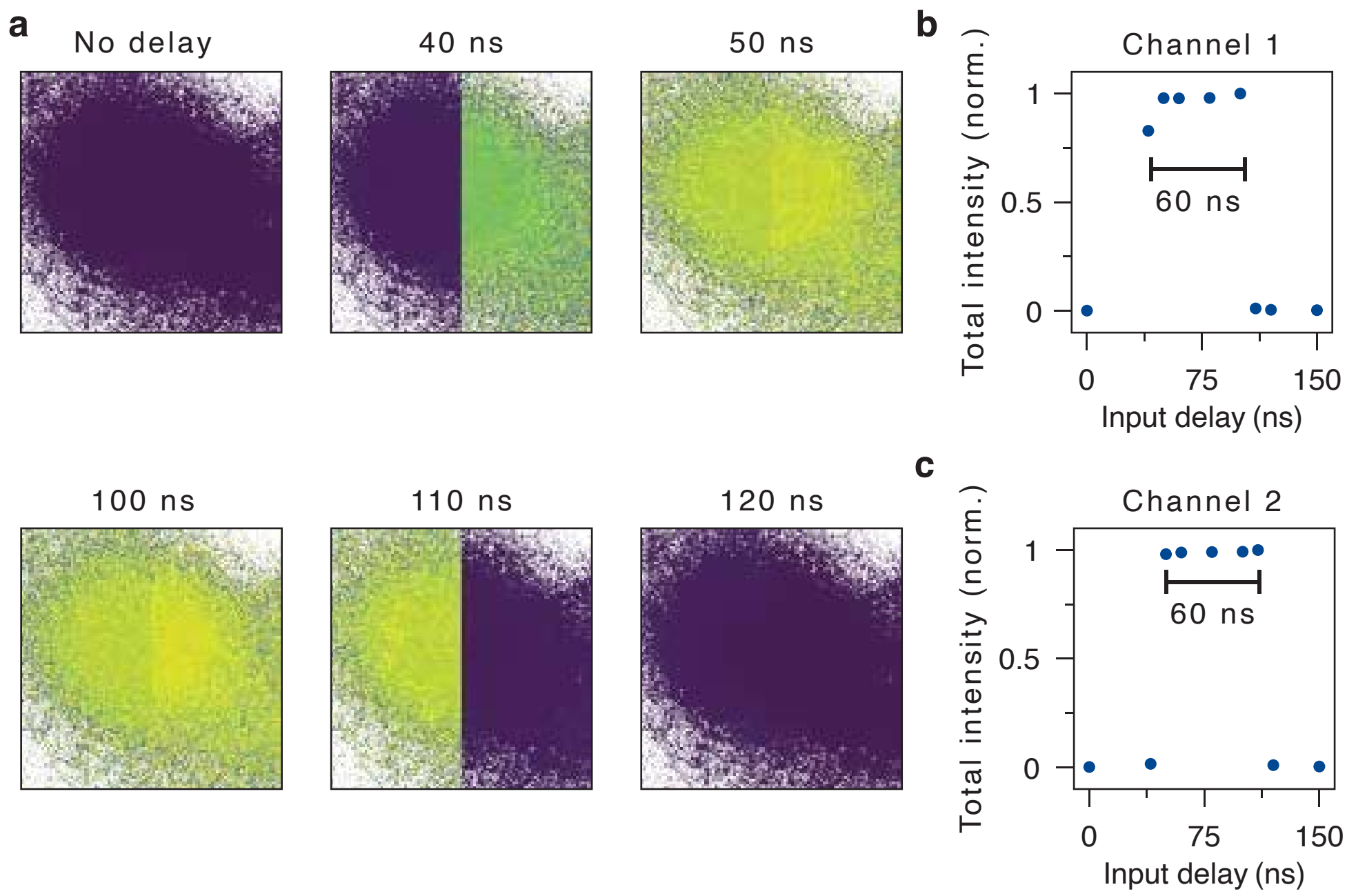

**Figure S5. SPAD delay calibration.** (a) Sum of 1000 gated frames recorded with a 20 ns integration window for different delays between the laser pulse and the start of the detector integration window. The images are normalized to the average intensity over all delay settings. The different panels correspond to detector delays of 0, 40, 50, 100, 110, and 120 ns. The images show an offset between the two detector halves, corresponding to the two detection channels used in the experiment. (b,c) Total normalized intensity as a function of the detector delay for channel 1 (b) and channel 2 (c). The laser pulse reaches channel 1 and channel 2 approximately 40 and 50 ns, respectively, after the start of the detector integration window. Although the integration window is 20 ns and the excitation pulse is much shorter, the detected signal extends over approximately 60 ns. This indicates a small uncertainty or spread in the effective detector triggering. The spatial maps in panel a show that this extended response is not caused by relative delays of individual pixels.

# S8 Cluster contributions and delayed-gate selection

## S8.1 Cluster contributions to the measured $g^{(2)}(0)$

The presence of unresolved clusters can distort the measured $g^{(2)}(0)$ values. If two or more quantum shells are present within the same diffraction-limited spot, photons emitted by different particles can contribute to the zero-delay coincidence peak. These inter-particle coincidences increase the $g^{(2)}(0)$, so a high measured value does not necessarily imply that the individual emitters have a high biexciton efficiency.[11] Because particles within a cluster can differ in brightness, the magnitude of this effect also depends on their relative intensities.

For a cluster containing $k$ emitters, we define the fractional brightness of emitter $i$ as

$$\alpha_i = \frac{I_i}{\sum_{j=1}^{k} I_j}. \tag{S30}$$

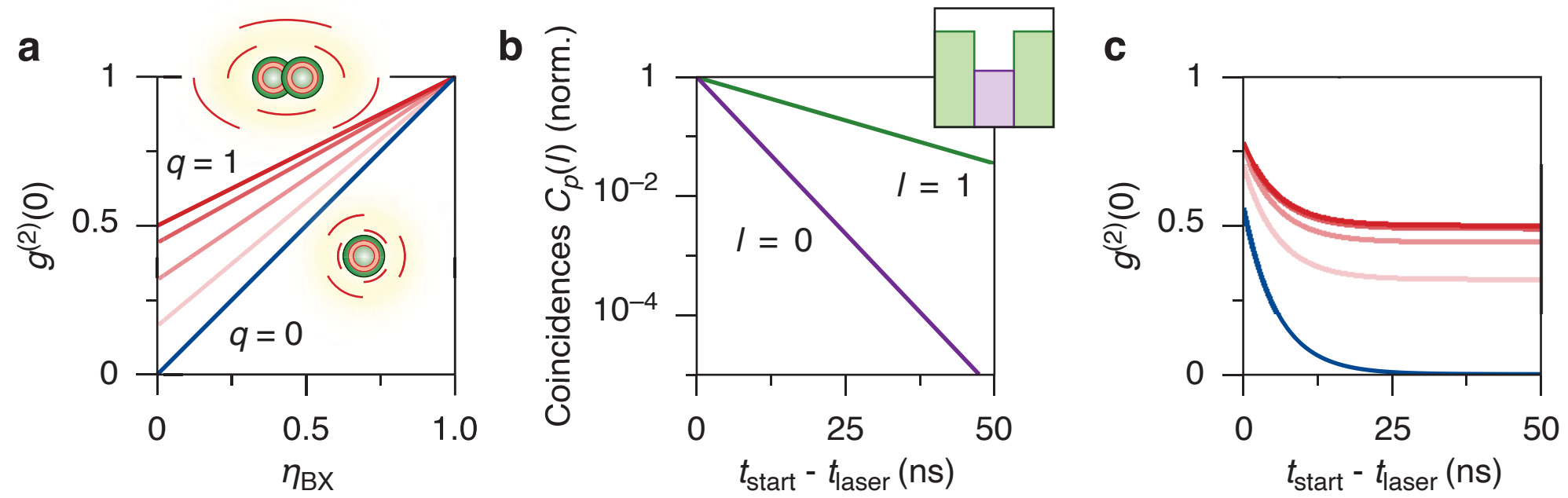

**Figure S6. Effect of unresolved clusters and delayed time gating on $g^{(2)}(0)$.** (a) $g^{(2)}(0)$ of a two-emitter cluster as a function of the intrinsic single-particle biexciton efficiency for different brightness ratios $q$. The blue line shows the single-emitter response, while the red lines show two-emitter clusters with $q = 1$, 0.75, 0.5, and 0.25. (b) Calculated gate dependence of the zero-delay biexciton-cascade contribution and the side-peak exciton contribution for a quantum shell with $\tau_X = 30$ ns and $\eta_{BX} = 0.55$. Delaying the start of the integration window suppresses the zero-delay biexciton contribution more strongly than the side-peak contribution. (c) Calculated normalized delayed-gate $g^{(2)}(0)$ for an individual emitter and for two-emitter clusters with different brightness ratios $q$. $g^{(2)}(0)$ decreases toward zero for an individual emitter, whereas two-emitter clusters retain a finite residual contribution from inter-particle exciton coincidences.

The photon correlation from the $k$-emitter cluster for a delay of $l \geq 1$ frame is

$$g_k^{(2)}(l) = \sum_{i,j} \alpha_i \alpha_j = 1. \tag{S31}$$

This summation describes coincidences between exciton photons emitted in different excitation cycles, so it includes terms with $i = j$. The summation adds up to 1 because of the definition of $\alpha_i$.

The zero-delay correlation of the cluster is

$$g_k^{(2)}(0) = \sum_{i=1}^{k} \alpha_i^2 g_i^{(2)}(0) + \sum_{i \neq j} \alpha_i \alpha_j. \tag{S32}$$

The first term describes biexciton–exciton cascades from individual emitters, weighted by their relative brightnesses. Here, $g_i^{(2)}(0)$ is the photon correlation of isolated emitter $i$. The second term describes inter-particle coincidences between photons emitted by different emitters in the same excitation cycle. We can use the identity

$$\sum_{i,j} \alpha_i \alpha_j = \sum_{i \neq j} \alpha_i \alpha_j + \sum_i \alpha_i^2 = 1 \tag{S33}$$

to write

$$g_k^{(2)}(0) = 1 + \sum_{i=1}^{k} \alpha_i^2 \left[ g_i^{(2)}(0) - 1 \right]. \tag{S34}$$

For a homogeneous cluster, with emitters of equal relative brightness and equal biexciton efficiency, $g_i^{(2)}(0) = g^{(2)}(0)$, we have

$$\left(\sum_{i=1}^{k} \alpha_i^2\right)_{\text{hom}} = k\left(\frac{1}{k}\right)^2 = \frac{1}{k}. \tag{S35}$$

The expected zero-delay correlation is then

$$g_k^{(2)}(0) = 1 + \frac{1}{k}\left[g^{(2)}(0) - 1\right]. \tag{S36}$$

To illustrate the consequences of Eq. S32, we consider a two-emitter cluster with unequal brightnesses (Figure S6a). We describe the relative brightness of the two emitters by

$$q = \frac{I_2}{I_1}, \qquad 0 < q \leq 1, \tag{S37}$$

where $I_1$ and $I_2$ are the intensities of the brighter and dimmer emitter, respectively. The limit $q \to 0$ corresponds to a strongly imbalanced cluster that behaves like a single emitter, whereas $q = 1$ corresponds to two equal-brightness emitters. The corresponding fractional brightnesses are

$$\alpha_1 = \frac{1}{1+q}, \qquad \alpha_2 = \frac{q}{1+q}. \tag{S38}$$

If both emitters have the same intrinsic biexciton efficiency, $g_i^{(2)}(0) = \eta_{\text{BX}}$, Eq. S32 reduces to

$$g_{\text{2em}}^{(2)}(0, q) = \frac{\eta_{\text{BX}}(1+q^2) + 2q}{(1+q)^2}. \tag{S39}$$

The first term in the numerator, $\eta_{\text{BX}}(1+q^2)$, describes biexciton coincidences from the two individual emitters, weighted by their relative brightnesses. The second term, $2q$, describes inter-particle coincidences between photons emitted by different emitters. As shown in Figure S6a, the cluster response approaches the single-emitter response $g^{(2)}(0) = \eta_{\text{BX}}$ in the limit $q \to 0$, because the dim emitter contributes negligibly. For equal brightness, $q = 1$, the inter-particle contribution is maximal and the expression reduces to

$$g_{\text{2em}}^{(2)}(0) = \frac{\eta_{\text{BX}} + 1}{2}. \tag{S40}$$

## S8.2 Time-gated $g^{(2)}$ for single emitters

To experimentally separate spots from individual emitters from spots containing clusters, we use delayed time gating.[11,12] This method relies on the different time dependence of biexciton-cascade coincidences and inter-particle exciton coincidences. For an individual quantum shell, zero-delay coincidences originate from the biexciton–exciton cascade. The biexciton photon is emitted first, followed by the exciton photon. Because biexciton emission is faster than exciton emission, shifting the start of the integration window to later times suppresses biexciton photons more strongly than exciton photons.

We denote the gate start time relative to the laser pulse as $t_0 = t_{\text{start}} - t_{\text{laser}}$. If an exciton photon is emitted, the probability that it falls within the detection time window scales as

$$p_{\mathrm{X}}(t_0) = \exp\left(-\frac{t_0}{\tau_{\mathrm{X}}}\right), \tag{S41}$$

where $\tau_{\mathrm{X}}$ is the exciton lifetime. Because the biexciton photon is emitted first in the cascade, the entire photon pair falls within the detection window if and only if the first photon does. The probability of this is

$$p_{\mathrm{BX}}(t_0) = \exp\left(-\frac{t_0}{\tau_{\mathrm{BX}}}\right), \tag{S42}$$

where $\tau_{\mathrm{BX}}$ is the biexciton lifetime.

For a single emitter, the zero-delay peak scales with the biexciton-pair detection probability, whereas the side peaks scale with the probability of detecting exciton photons in two independent excitation cycles. We therefore write the delayed-gate response of a single emitter as

$$\tilde{g}^{(2)}(0, t_0) = \eta_{\mathrm{BX}} \frac{p_{\mathrm{BX}}(t_0)}{p_{\mathrm{X}}(t_0)^2}. \tag{S43}$$

The normalization is chosen such that

$$\tilde{g}^{(2)}(0, 0) = \eta_{\mathrm{BX}}. \tag{S44}$$

Because $\tau_{\mathrm{BX}} < \tau_{\mathrm{X}}$, the biexciton contribution in the numerator decreases more rapidly with $t_0$ than the side-peak normalization in the denominator. As a result, the delayed-gate $\tilde{g}^{(2)}(0, t_0)$ of an individual emitter decreases toward zero as the gate is shifted to later times (Figure S6b,c).

## S8.3 Time-gated $g^{(2)}$ for clusters

For a cluster, zero-delay coincidences can additionally arise from independent exciton photons emitted by different particles. These inter-particle coincidences are not biexciton-cascade events and are therefore not suppressed in the same way by delayed gating. Using Eq. S34, the delayed-gate response of a $k$-emitter cluster can be written as

$$\tilde{g}_k^{(2)}(0, t_0) = 1 + \sum_{i=1}^{k} \alpha_i^2 \left[\tilde{g}_i^{(2)}(0, t_0) - 1\right]. \tag{S45}$$

With a detection time gate starting at a sufficiently large $t_0 \gg \tau_{\mathrm{BX}}$, the single-particle biexciton contribution is suppressed, so that $\tilde{g}_i^{(2)}(0, t_0)$ approaches zero. The residual cluster contribution then becomes

$$\tilde{g}_k^{(2)}(0, t_0 \gg \tau_{\mathrm{BX}}) = 1 - \sum_{i=1}^{k} \alpha_i^2. \tag{S46}$$

For a two-emitter cluster this residual contribution is

$$\tilde{g}_{2\mathrm{em}}^{(2)}(0, t_0 \gg \tau_{\mathrm{BX}}) = \frac{2q}{(1+q)^2}. \tag{S47}$$

For equal brightness, $q = 1$, the delayed-gate cluster response therefore approaches 0.5. For unequal brightnesses, the residual cluster value shifts to lower $g^{(2)}(0)$, as shown in Figure S6c.

To estimate the effect of brightness imbalance in our measurements, we used the experimentally measured single-particle brightness distribution. As described below, random sampling of this distribution gives an effective two-emitter brightness ratio of $q_{\text{eff}} = 0.483$. For this value, Eq. S47 gives

$$\frac{2q_{\text{eff}}}{(1 + q_{\text{eff}})^2} = 0.44. \tag{S48}$$

We therefore used

$$g^{(2)}_{\text{delayed}}(0) < 0.4 \tag{S49}$$

as a conservative criterion for selecting spots with a delayed-gate response consistent with individual emitters. The remaining spots were treated as cluster-like. The delayed-gate selection therefore removes most equal-brightness and moderately imbalanced clusters while retaining the main single-emitter population.

## S8.4 Predicting cluster peak positions in the raw histogram

In Figure 5f of the main text, we used the mean single-quantum-shell value $\langle g^{(2)}(0)\rangle_1$ to predict the expected mean $\langle g^{(2)}(0)\rangle_k$ for unresolved clusters containing $k = 2, 3, 4$, or 5 emitters. We first fitted the selected single-emitter distribution with a Gaussian model and used the fitted mean $\langle g^{(2)}(0)\rangle_1$ as the average single-particle biexciton efficiency.

The quantum shells in our experiments vary in brightness. The brightness variations affect $g^{(2)}_k(0)$ through $\sum_i \alpha_i^2$ in Eq. S34. We use the experimental brightness variations to simulate $\sum_i \alpha_i^2$ for clusters. We first note that the experimental brightness distribution contains contributions from:

- Intrinsic differences in brightness between quantum shells. This is what we want to consider for our calculations.
- Clustering. We want to exclude the contribution of clusters of two or more quantum shells in our calculations, and therefore excluded spots with $g^{(2)}(0) > 0.65$ in the undelayed measurement or with $g^{(2)}_{\text{delayed}}(0) > 0.4$ in the time-gated measurement with $t_{\text{start}} = 120$ ns.
- Variations in laser excitation intensity over the sample surface. To correct for this, we normalize the count rate of a spot by the local background-fluorescence intensity, assuming that the local background fluorescence scales with local laser intensity.

These procedures yield a set of experimental brightness levels $I_i$, which we interpret as the intrinsic brightness distribution of the quantum-shell sample.

Next, we use this brightness distribution to simulate clusters of $k$ quantum shells and calculate the average $\langle g^{(2)}(0)\rangle_k$. Specifically, we evaluate the average $\langle \sum_{i=1}^{k} \alpha_i^2 \rangle$ numerically by drawing 50,000 sets of $k$ emitters from the brightness distribution. Neglecting the correlation between brightness and biexciton efficiency, which is present in reality

(Figure S7), we insert $\langle\sum_{i=1}^{k}\alpha_i^2\rangle$ into Eq. S34 and obtain

$$\langle g^{(2)}(0)\rangle_k = 1 + \left\langle \sum_{i=1}^{k} \alpha_i^2 \right\rangle \left[\langle g^{(2)}(0)\rangle_1 - 1\right]. \tag{S50}$$

The values of $\langle g^{(2)}(0)\rangle_k$ for $k = 1, 2, \ldots, 5$ are shown as arrows in Figure 5f.

For the two-emitter case, the same sampling procedure also gives an effective brightness moment,

$$s_{2,\mathrm{eff}} = \left\langle \alpha_1^2 + \alpha_2^2 \right\rangle. \tag{S51}$$

We convert this value into an effective two-emitter brightness ratio using

$$q_{\mathrm{eff}} = \frac{s_{2,\mathrm{eff}} - \sqrt{2s_{2,\mathrm{eff}} - 1}}{1 - s_{2,\mathrm{eff}}} = 0.483. \tag{S52}$$

This $q_{\mathrm{eff}}$ is not the arithmetic mean of the sampled intensity ratios. Instead, it is the two-emitter brightness ratio that gives the same average cluster contribution as the experimentally measured brightness distribution.

# S9 Brightness-dependent biexciton efficiency

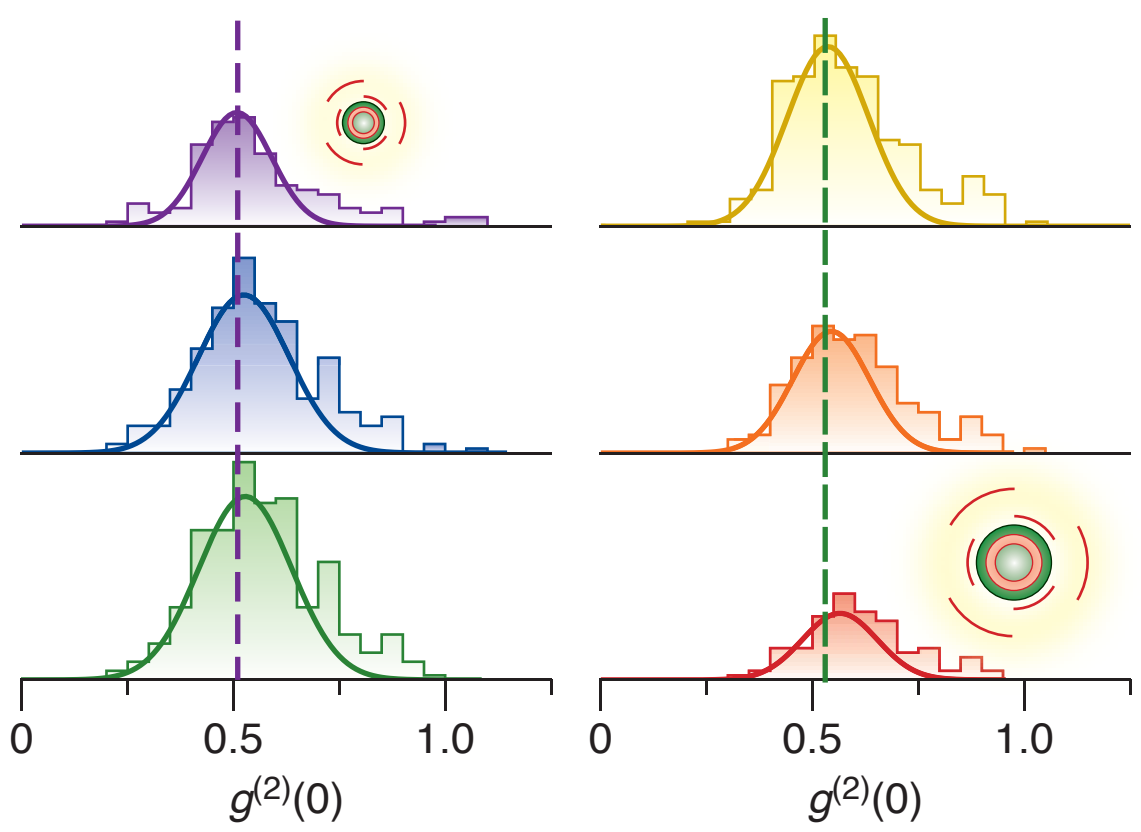

**Figure S7. Brightness-dependent $g^{(2)}(0)$ distributions.** Selected single-emitter-enriched $g^{(2)}(0)$ distributions grouped by relative fluorescence brightness. The relative brightness was calculated from the summed intensity in the two matched reference images after background correction and normalization. Particles were grouped into overlapping brightness bins with a width of 5000 intensity units; the lower bin edge was shifted from 0 to 15000 in steps of 2500. The color scale runs from lower relative brightness (purple) to higher relative brightness (red), corresponding to the brightness bins used in Figure 5g of the main text. Solid lines show Gaussian fits to each distribution. The fitted mean $g^{(2)}(0)$, interpreted as the mean biexciton efficiency, increases with particle brightness.